\documentclass[prd,superscriptaddress,twocolumn,nofootinbib,showpacs]{revtex4}

\usepackage{amsfonts,amssymb,amsmath}
\usepackage{mathrsfs}
\usepackage{color}
\usepackage{graphicx}% Include figure files
\usepackage{dcolumn}% Align table columns on decimal point
\usepackage{multirow}
\usepackage{bm}% bold math
\newcommand{\B}[1]{{\mathbb #1}}
\newcommand{\C}[1]{{\mathcal #1}}

\newcommand{\BS}[1]{{\boldsymbol #1}}

\newcommand{\beq}{\begin{equation}}
\newcommand{\eeq}{\end{equation}}
\newcommand{\bea}{\begin{eqnarray}}
\newcommand{\eea}{\end{eqnarray}}
\newcommand{\nn}{\nonumber}

\newcommand{\half}{\frac 12}
\newcommand{\third}{\frac 13}
\newcommand{\quarter}{\frac 14}

\newcommand{\sixth}{\frac 16}

\newcommand{\Slash}[1]{{\ooalign{\hfil#1\hfil\crcr\raise.167ex\hbox{/}}}}

\begin{document}

%%%%%%%%%%%%%%%%%%%%%%%%%%%%%%%%%%%%%%%%%%%%%%%%

%\preprint{arXiv:yymm.nnnn[hep-**]}

%%%%%%%%%%%%%%%%%%%%%%%%%%%%%%%%%%%%%%%%%%%%%%%%

\title{Testing supersymmetric Higgs inflation with non-Gaussianity}

\author{Shinsuke Kawai}
\email{kawai(AT)skku.edu}
\affiliation{Department of Physics, 
Sungkyunkwan University,
Suwon 440-746, Korea}
\author{Jinsu Kim}
\email{kimjinsu(AT)skku.edu}
\affiliation{Department of Physics, 
Sungkyunkwan University,
Suwon 440-746, Korea}

%\date{\today}
\date{February 13, 2015}

%%%%%%%%%%%%%%%%%%%%%%%%%%%%%%%%%%%%%%%%%%%%
%%%%%%%%%%%%%%%%%%%%%%%%%%%%%%%%%%%%%%%%%%%%

%%%%% Abstract %%%%%

\begin{abstract}
We investigate multifield signatures of the nonminimally coupled supersymmetric
Higgs inflation-type cosmological scenario, focusing on 
the two-field Higgs-lepton inflation model as a concrete example.
This type of inflationary model is realized in a theory beyond the Standard Model embedded in supergravity with a noncanonical K\"{a}hler potential.
We employ the backward $\delta N$ formalism to compute cosmological observables, including the scalar and tensor power spectra, the spectral indices, the tensor-to-scalar ratio, and the local-type nonlinearity parameter.
The trajectory of the inflaton is controlled by the initial conditions of the inflaton as well as by the coefficients in the K\"{a}hler potential.
We analyze the bispectrum of the primordial fluctuations when the inflaton trajectory deviates from a straight line and obtain constraints on the noncanonical terms of the K\"{a}hler potential using the Planck satellite data.
Our analysis represents a concrete particle phenomenology-based case study of inflation in which primordial non-Gaussianities can reveal aspects of supergravity.
\end{abstract}

%%%%%%%%%%%%%%%%%%%%%%%%%%%%%%%%%%%%%%%%%%%%
%%%%%%%%%%%%%%%%%%%%%%%%%%%%%%%%%%%%%%%%%%%%

\pacs{12.60.Jv, 04.65.+e, 98.80.Cq, 98.70.Vc}
\keywords{Supersymmetric models, Supergravity, Inflation, Cosmic microwave background}
\maketitle

%%%%%%%%%%%%%%%%%%%%%%%%%%%%%%%%%%%%%%%%%%%%
%%%%%%%%%%%%%%%%%%%%%%%%%%%%%%%%%%%%%%%%%%%%

%%%%% Body of the Paper %%%%%

%%%%%%%%%%%%%%%%%%%%%%%%%%%%%%%%%%%%%%%%%%%%
%%%%%%%%%%%%%%%%%%%%%%%%%%%%%%%%%%%%%%%%%%%%
\section{Introduction}\label{sec:Intro}

In cosmology, the precision of measurements has dramatically improved in the last decade or so.
The recent Planck satellite experiments of the cosmic microwave background (CMB), for example, indicate that the scalar spectral index $n_s$, the tensor-to-scalar ratio $r$, and the local-type nonlinearity parameter $f_{\rm NL}^{\text{local}}$ are in the following 
windows \cite{Ade:2013zuv,Ade:2013uln,Ade:2013ydc}:
\begin{align}
n_{s} = &\ 0.9603\pm 0.0073 & (68\%\text{ C.L.}), \cr
r < &\ 0.12 & (95\%\text{ C.L.}),\cr
f^{\text{local}}_{\rm{NL}} = &\ 2.7\pm 5.8 & (68\%\text{ C.L.}).
\label{eqn:planck}
\end{align}
Eventually, these data are to be accounted for by a model of the Universe based on, ideally, a well-motivated theory of particle physics.
The leading account of the early Universe in agreement with present observational data
is inflationary cosmology, which emerged as a solution to the flatness, horizon, and monopole problems of
the standard big bang cosmology. 
Currently, inflationary model building is somewhat postmodernistic---there is a plethora of toy models inspired by string theory and M theory, among others, 
and many of them can be adjusted to fit the data.
Future observation could change this situation, however, as measurements with increasing accuracy are expected to put many models under pressure.

% Supersymmetric models
To build a realistic cosmological scenario beyond inflationary toy models, the supersymmetric extension of the Standard Model provides a technically natural and phenomenologically well-motivated framework.
A consistent scenario of cosmology needs to be compatible with physics at low energies, including particle phenomenology at collider scales, and thus must incorporate the Standard Model in some form.
Moreover, if the energy scale of inflation turns out to be as high as $H\approx 10^{14}$ GeV 
($H$ is the Hubble parameter) as implied\footnote{
Presuming that the observed B-mode polarization results from the primordial tensor mode fluctuations.
See also Refs. \cite{Mortonson:2014bja,Flauger:2014qra,Adam:2014oea}.}
by the BICEP2 experiments \cite{Ade:2014xna}, 
it is plausible that supersymmetry plays some role in the physics of inflation.
Recently there has been a keen interest in the Standard Model Higgs inflation model \cite{CervantesCota:1995tz,Bezrukov:2007ep}, in which the gravitationally coupled Higgs field is identified as the inflaton.
A supersymmetric version of the Higgs inflation model was implemented first in the next-to-minimal supersymmetric Standard Model (NMSSM) 
\cite{Einhorn:2009bh,Ferrara:2010yw,Ferrara:2010in,Einhorn:2012ih}.
Subsequently, various other models---based on the supersymmetric Pati--Salam model \cite{Pallis:2011gr}, the supersymmetric grand unified theory \cite{Arai:2011nq,Ellis:2014dxa}, 
the supersymmetric B-L model \cite{Arai:2013vaa}, and the 
supersymmetric seesaw model \cite{Arai:2011aa,Arai:2012em,Kawai:2014doa}---were proposed.
In contrast to the Standard Model Higgs inflation model, these supersymmetric models necessarily involve multiple scalar fields participating in the dynamics of inflation.
The effects of multiple fields, so far, have not been studied in full detail, due to the complexities
pertaining to the larger degrees of freedom.

% Purpose
In this paper, we discuss non-Gaussianities of the primordial fluctuations in these supersymmetric Higgs inflation models.
It is well known that single-field inflation typically predicts primordial fluctuations of the Gaussian spectrum;
hence, detection of sizeable non-Gaussianities would be strong evidence for multifield inflation.
Since present observation of cosmological parameters is all consistent with the prediction of single-field inflation \cite{Ade:2013uln}, we shall take a modest approach and start from a single-field limit, that is,
inflation with a straight inflaton trajectory.
We then analyze how the prediction for the bispectrum changes as the trajectory deviates from a straight line.
For the sake of concreteness, we consider the inflationary model based on the supersymmetric
seesaw model, which is dubbed the supersymmetric Higgs-lepton inflation (HLI) model
\cite{Arai:2011aa,Arai:2012em,Kawai:2014doa}.
Also, we focus on the two-field case for simplicity.
To compute the fluctuation spectrum of the inflationary model,
we use the backward formulation 
\cite{Yokoyama:2007dw, Yokoyama:2007uu, Yokoyama:2008by} of the $\delta N$-formalism
\cite{Sasaki:1995aw,Nakamura:1996da,Sasaki:1998ug,Lyth:2004gb,Lyth:2005fi,Seery:2005gb}.
%It has been shown that the $\delta N$-formalism is physically equivalent to the covariant formalism
%\cite{Naruko:2012um,Suyama:2012wi}.
We find by numerical computations that the bispectrum of the inflationary model is susceptible to a change of the inflaton trajectory,
a fact known in generic cases; see, e.g., Refs. \cite{Peterson:2010np, Peterson:2010mv, Peterson:2011yt, Elliston:2012ab}.
Since the shape of the trajectory depends on a parameter of the K\"{a}hler potential in the class
of inflationary models we consider,
constraints on the K\"{a}hler potential are obtained from the experimental bounds of 
non-Gaussianities \eqref{eqn:planck}.
While the details can be model dependent, the generic features of the outcome should be common
in other similar models.
To illustrate another example of supersymmetric Higgs inflation, we comment on the NMSSM-based model in Appendix \ref{apdx:NMSSM}.

%covariant formalism 
%\cite{Langlois:2005ii, Langlois:2005qp, Langlois:2006vv, Langlois:2008vk, RenauxPetel:2008gi}, 
%standard perturbative approach
%\cite{Kaiser:2013sna, Acquaviva:2002ud, Malik:2003mv, Lyth:2005du}, 

Non-Gaussianities have been studied extensively in various multifield inflationary models.
The possibility of generating large local-type non-Gaussianities is pointed out in Ref. \cite{Byrnes:2008wi},
and the conditions for it are studied in Refs. \cite{Wang:2010si,Byrnes:2009qy}.
The literature on inflationary models with nontrivial field space resulting from nonminimal coupling includes Refs. \cite{Elliston:2012ab,Kaiser:2012ak,Kaiser:2013sna}.
Multifield analyses of supergravity-based inflationary toy models similar to ours in spirit include Refs. 
\cite{Linde:2012bt,Choudhury:2014uxa,Ellis:2014opa}.

The rest of this paper is organized as follows. 
In Sec. \ref{sec:HLI}, we illustrate the HLI model, which is our main focus.
In Sec. \ref{sec:deltaN}, we give a brief review of the backward $\delta N$ formalism and define quantities describing cosmological observables.
The numerical results are shown in Sec. \ref{sec:result}, and observational constraints on the parameter space are also discussed there.
We conclude in Sec. \ref{sec:concl} with comments.
Some formulas of the $\delta N$ formalism are collected in Appendix \ref{apdx:deltaN}, and 
the NMSSM-based supersymmetric Higgs inflation model is described briefly 
in Appendix \ref{apdx:NMSSM}.

%%%%%%%%%%%%%%%%%%%%%%%%%%%%%%%%%%%%%%%%%%%%
%%%%%%%%%%%%%%%%%%%%%%%%%%%%%%%%%%%%%%%%%%%%

%%%%%%%%%%%%%%%%%%%%%%%%%%%%%%%%%%%%%%%%%%%%
%%%%%%%%%%%%%%%%%%%%%%%%%%%%%%%%%%%%%%%%%%%%
\section{Inflationary model}\label{sec:HLI}

In this paper, we consider a model of inflation described by the Lagrangian density\footnote{
Our conventions are
$g_{\mu\nu}=(-,+,+,+)$,
%\Gamma^\lambda_{\mu\nu}=\half g^{\lambda\rho}
%(g_{\rho\mu,\nu}+g_{\rho\nu,\mu}-g_{\mu\nu,\rho}),\nn\\
${\C R}^\lambda{}_{\mu\rho\nu}=\Gamma^\lambda_{\mu\nu,\rho}-\Gamma^\lambda_{\mu\rho,\nu}
+\Gamma^\lambda_{\kappa\rho}\Gamma^\kappa_{\mu\nu}-\Gamma^\lambda_{\kappa\nu}\Gamma^\kappa_{\mu\rho}$,
${\C R}_{\mu\nu}={\C R}^\rho{}_{\mu\rho\nu}$,
and
${\C R}=g^{\mu\nu}{\C R}_{\mu\nu}$
for the spacetime and similarly for the field space except that the metric is $G_{ab}=(+,+)$.
The reduced Planck mass $M_{\rm P}=1/\sqrt{8\pi G_{\rm N}}=2.436 \times 10^{18}$ GeV is
set to unity unless otherwise indicated.}
\begin{align}
\mathcal{L}=\sqrt{-g}\left[\frac{1}{2}{\C R}-\frac{1}{2}G_{IJ}g^{\mu\nu}\partial_{\mu}\phi^{I}\partial_{\nu}\phi^{J}-V(\phi^I)\right]\,,
\label{eqn:Lag}
\end{align}
where $g_{\mu\nu}$ is the spacetime metric
(we consider the flat Friedmann--Lema\^{i}tre--Robertson--Walker background metric), 
${\C R}$ is the Ricci scalar of the spacetime, and $g\equiv\det g_{\mu\nu}$.
We have two real scalar fields $\phi^{1}\equiv s$ and $\phi^{2}\equiv h$.
The indices are $\mu,\nu,\cdots = 0,1,2,3$ and $I,J,\cdots = 1,2$.
We will be interested in the special form of the field space metric $G_{IJ}$ given by
\begin{align}
G_{ss} &= \frac{\frac{1}{12}\upsilon s^{4}+(1-2\upsilon s^{2})(1+\xi h^{2})}{\Phi^{2}}\,,\nonumber\\
G_{sh}&=G_{hs}=-\frac{\xi h s (1-\upsilon s^{2})}{\Phi^{2}} \,,\nonumber\\
G_{hh} &= \frac{6\xi^{2}h^{2}+\Phi}{\Phi^{2}}\,,
\label{eqn:FSmetric}
\end{align}
where $\xi$, $\upsilon$ (Greek letter upsilon) are real parameters and
\begin{align}
\Phi\equiv 1-\frac{1}{6}s^{2}+\frac{1}{12}\upsilon s^{4}+\xi h^{2}.
\label{eqn:Phi}
\end{align}
%
%\begin{align}
%G_{11} &= \frac{\frac{1}{12}\upsilon s^{4}+(1-2\upsilon s^{2})(1+\xi h^{2})}{\Phi^{2}}\,,\nonumber\\
%G_{12}&=G_{21}=-\frac{\xi h s (1-\upsilon s^{2})}{\Phi^{2}}\,,\nonumber\\
%G_{22}&=\frac{6\xi^{2}h^{2}+\Phi}{\Phi^{2}}\,.
%\end{align}
%
The Christoffel symbol on the field space is
\begin{align}
\Gamma^{1}_{11}&=\frac{\upsilon s\big[s^{2}-12\{1+(1+6\xi)\xi h^{2}\}\big]}{6C}
+\frac{s(1-\upsilon s^{2})}{3\Phi}\,,\nonumber\\
\Gamma^{1}_{12}&=-\frac{\xi h}{\Phi}\,,\quad
\Gamma^{1}_{22}=-\frac{(1+6\xi)(1-\upsilon s^{2})s}{6C}\,,\nonumber\\
\Gamma^{2}_{11}&=-\frac{\upsilon s^{2}\xi h}{C},\qquad
\Gamma^{2}_{12}=\frac{(1-\upsilon s^{2})s}{6\Phi}\,,\\
\Gamma^{2}_{22}&=\frac{12(1-\xi h^{2})+\upsilon s^{4}-2s^{2}}{12h\Phi}
-\frac{12+(s^{2}-24)\upsilon s^{2}}{12hC}\,,\nn
\label{eqn:Gamma}
\end{align}
where
\begin{align}
C&\equiv \Phi^{3} \mathrm{det} G_{IJ}\,\nonumber\\
&=1-2\upsilon s^{2}+(1+6\xi)(1-2\upsilon s^{2})\xi h^{2}+\frac{1}{12}\upsilon s^{4}\,.
\end{align}
The scalar curvature of the field space is
\begin{align}
R=-\frac{1}{3}-\frac{(1+6\xi)\upsilon s^{2}\Phi^{2}}{3C^{2}}\,.
\end{align}
Note that the Riemann curvature is written by the scalar curvature as 
$R^I{}_{JKL}=\half R(\delta^I_K G_{JL}-\delta^I_L G_{JK})$ in two dimensions.

The two-field Lagrangian \eqref{eqn:Lag} with the field space metric \eqref{eqn:FSmetric} 
is obtained from supergravity with a particular type of (noncanonical) K\"{a}hler potential.
This class of cosmological scenario includes those based on the NMSSM 
\cite{Einhorn:2009bh,Ferrara:2010yw,Ferrara:2010in,Einhorn:2012ih},
the supersymmetric Pati--Salam model \cite{Pallis:2011gr}, the supersymmetric grand unified theory \cite{Arai:2011nq}, and the supersymmetric seesaw model \cite{Arai:2011aa,Arai:2012em,Kawai:2014doa}.
The form of the potential $V(\phi^I)$ depends on details of each phenomenological setup.
In this paper, we focus on the HLI model based on the supersymmetric seesaw.
Below in this section, we review the construction of this model 
\cite{Arai:2011aa,Arai:2012em,Kawai:2014doa}.
For comparison, we sketch the model based on the NMSSM in Appendix \ref{apdx:NMSSM}.

%%%%%%%%%%%%%%%%%%%%%%%%%%%%%%%%%%%%%%%%%%%%
\subsection{Supersymmetric seesaw model}
%%%%%%%%%%%%%%%%%%%%%%%%%%%%%%%%%%%%%%%%%%%%

%%%%%%%%%%%%%%%%%%%%%%%%%%%%%%%%%%%%%%%%%%
%\begin{center}
\begin{figure*}
%\begin{eqnarray*}
%\begin{array}{ccc}
\includegraphics[width=59mm]{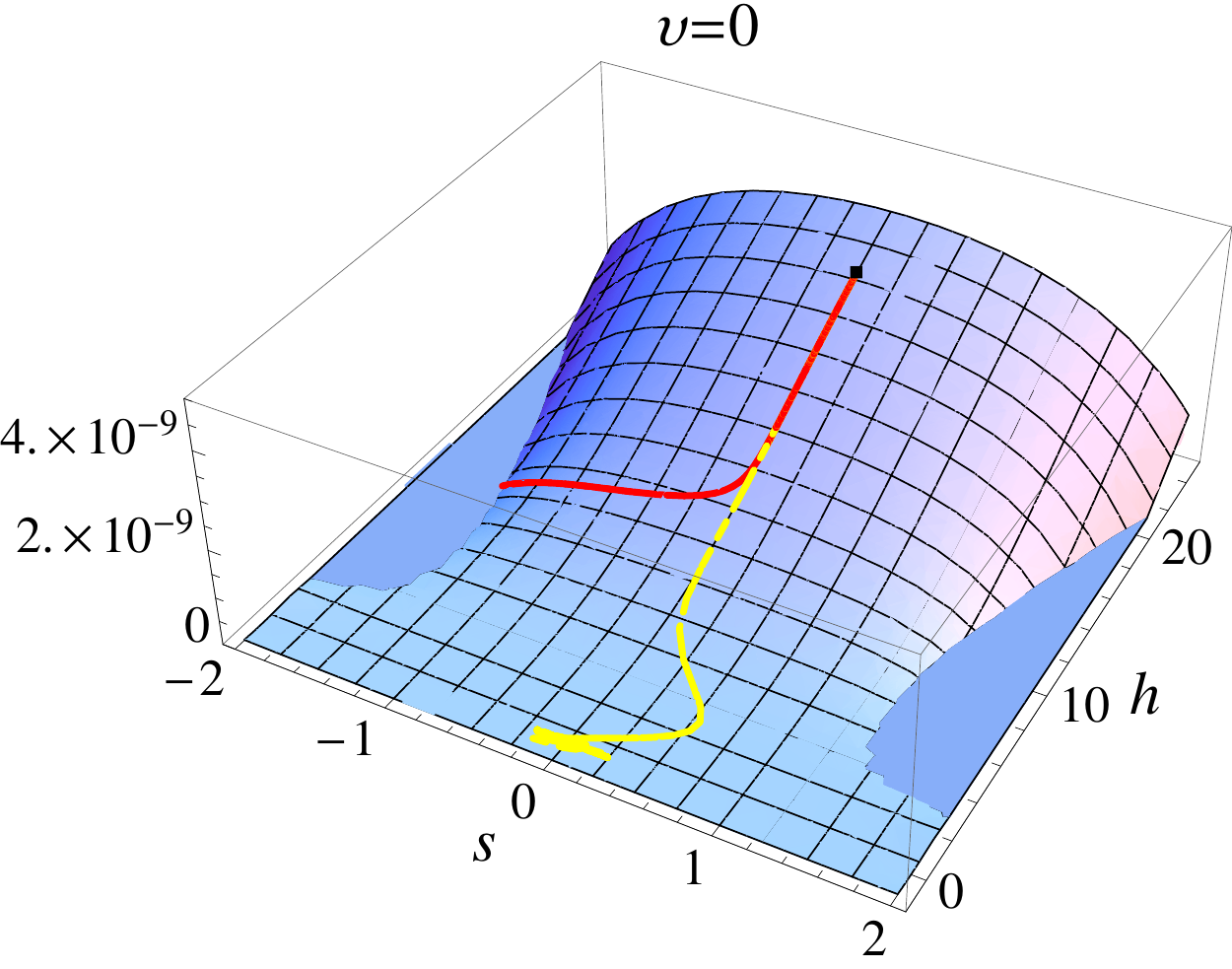}
%\includegraphics[width=59mm]{Fig1a_r.eps}
%&&
\includegraphics[width=59mm]{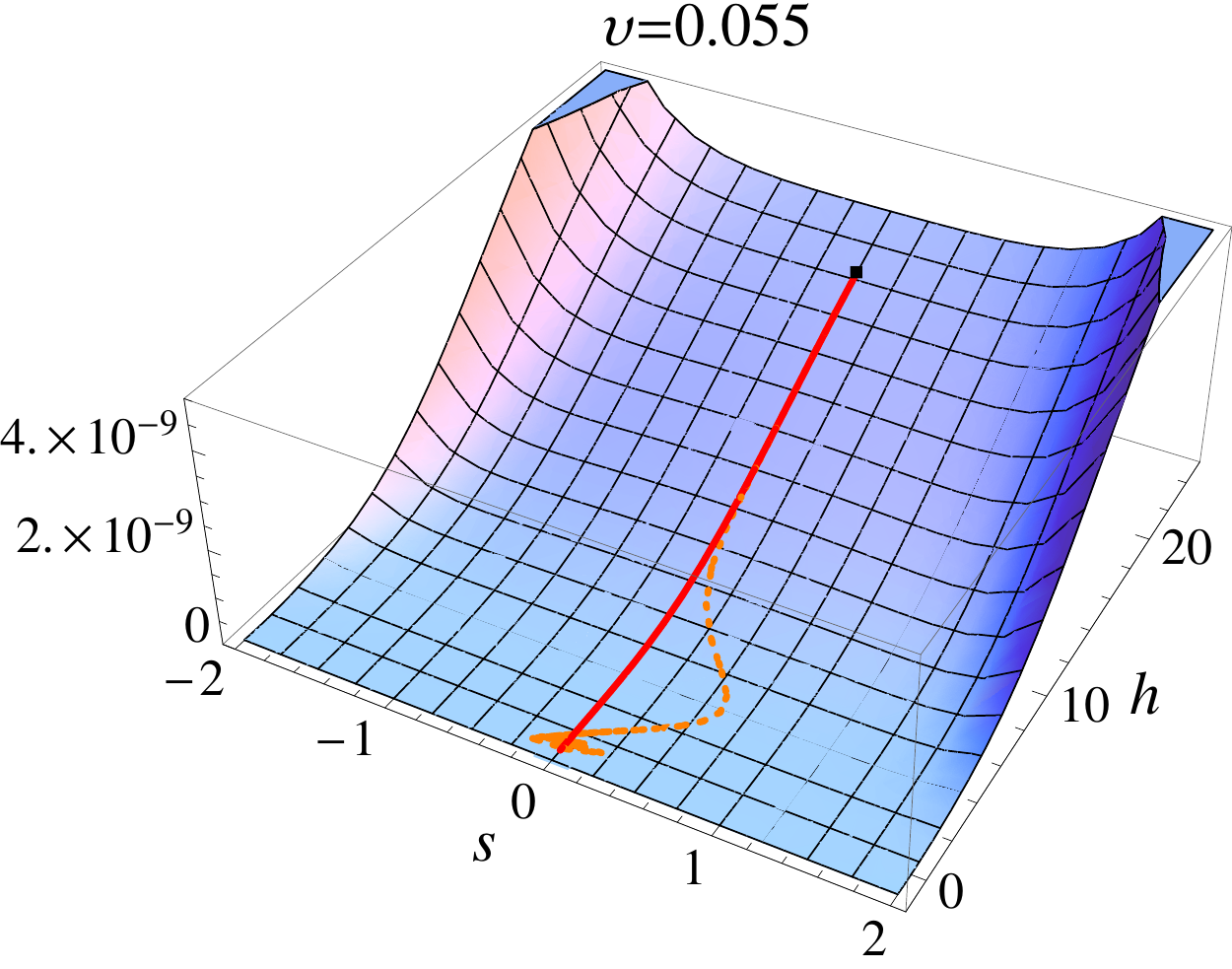}
%\includegraphics[width=59mm]{Fig1b_r.eps}
%&&
\includegraphics[width=59mm]{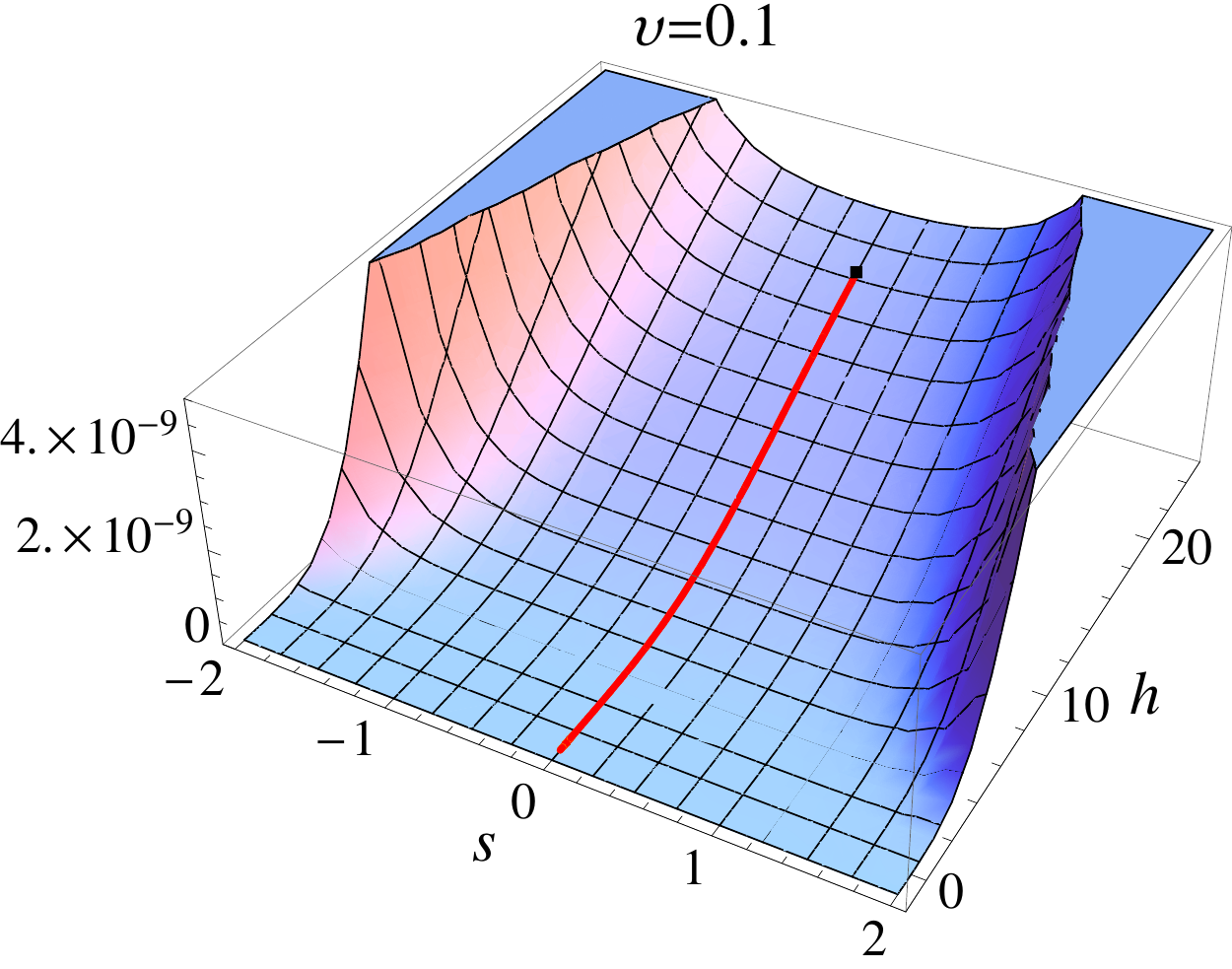}
%\includegraphics[width=59mm]{Fig1c_r.eps}
%\end{array}
%\end{eqnarray*}
\caption{\label{fig:PotView}
The shape of the scalar potential $V(\phi^I)$ for $M=1$ TeV and $\upsilon=0$ (left), 
$\upsilon=0.055$ (center), and $\upsilon=0.1$ (right).
The parameter $\xi$ is fixed to $\xi= 3.696\times 10^{-3}$ by the condition that
in the single-field limit the amplitude of the curvature perturbation corresponding to $N_e=60$ 
$e$-folds is Planck normalized $A_s=2.215\times 10^{-9}$ \cite{Ade:2013uln}.
The red curves are the inflaton trajectories with initial conditions 
$s_{\rm init}=0$, $\dot s_{\rm init}=0$ at $h=h_{\rm init}=21.99$
(this value of $h_{\rm init}$ corresponds to $N_e=60$ $e$-folds in the single-field limit);
the initial value for $\dot h$ is determined by the slow-roll equation of motion.
On each panel, the point $(s, h) = (0, 21.99)$ is marked with a black dot.
On the left panel ($\upsilon=0$), the flat regions on the sides represent negative $V(\phi^I)$, which are considered unphysical.
For small values of $\upsilon$, a trajectory can reach the supersymmetric vacuum $(s, h) = (0, 0)$ only when the initial conditions
are fine-tuned ($s_{\rm init} = 1.617\times 10^{-11}$, $\dot s_{\rm init}=0$ for the yellow dashed curve).
For generic initial conditions, the inflaton will fall into either of the $V(\phi^I)<0$ regions
(so does the red curve in the case of $s_{\rm init}=0$, $\dot s_{\rm init}=0$).
When $\upsilon=0.055$ (center), the potential is stabilized in the $s$-field direction.
The orange dotted curve that makes a mild turn 
corresponds to $s_{\rm init}=1.0\times 10^{-5}$, $\dot s_{\rm init}=0$.
When $\upsilon=0.1$ (right), the trajectories are more convergent. 
Two trajectories [initial conditions $(s_{\rm init},\dot s_{\rm init})=(0,0)$
and $(1.0 \times 10^{-5}, 0)$] are shown, but they are almost indistinguishable.
}
\end{figure*}
%\end{center}
%%%%%%%%%%%%%%%%%%%%%%%%%%%%%%%%%%%%%%%%%%

The supersymmetric seesaw model is an extension of the minimal supersymmetric Standard Model (MSSM) by adding a right-handed neutrino superfield $N_{R}^c$.
Its simplest version is described by the superpotential
\begin{align}
W = W_{\rm{MSSM}}+\frac{1}{2}M N_{R}^{c}N_{R}^{c}
+y_{D}N_{R}^{c}LH_{u}
\label{eqn:Wseesaw}
\,,
\end{align}
where $y_{D}$ is the Dirac Yukawa coupling, 
$M$ is the seesaw mass parameter, and
\begin{align}
W_{\rm{MSSM}}=\mu H_{u}H_{d}+y_{u}u^{c}QH_{u}+y_{d}d^{c}QH_{d}+y_{e}e^{c}LH_{d}
\label{eqn:WMSSM}
\,,
\end{align}
with the MSSM superfields $Q$, $u^c$, $d^c$, $L$, $e^c$, $H_{u}$, and $H_{d}$. 
In Eq. \eqref{eqn:WMSSM}, $\mu$ is the MSSM $\mu$ parameter, and $y_{u},~y_{d}$, and $y_{e}$ are the Yukawa couplings.
Assuming odd R parity for $N_{R}^c$, the superpotential \eqref{eqn:Wseesaw} preserves the R parity.
For generation of the small nonvanishing (left-handed) neutrino masses by the seesaw mechanism
\cite{seesaw},
%\cite{Minkowski:1977sc, Yanagida:1979as, GellMann:1980vs, Mohapatra:1979ia}, 
the Dirac Yukawa coupling $y_{D}$ and the right-handed neutrino mass $M$ in Eq. \eqref{eqn:Wseesaw} must satisfy the seesaw relation 
\begin{align}
m_{\nu} = \frac{y_{D}^{2} \langle H_{u}\rangle^{2}}{M}\,,
\end{align}
where $m_{\nu}$ is the left-handed neutrino mass and $\langle H_{u}\rangle\approx 174\,\rm{GeV}$
is the Higgs vacuum expectation value at low energies. 
Evaluating the neutrino mass by $m_{\nu}^{2} = \Delta m_{32}^{2} \approx 2.44\times10^{-3}\,\rm{eV^{2}}$ 
\cite{Agashe:2014kda}%PDG2014
, we find
\begin{align}
y_{D} = \left(\frac{M}{6.13\times 10^{14}\,\rm{GeV}}\right)^{1/2}\,.
\label{eqn:seesaw}
\end{align}
%

%%%%%%%%%%%%%%%%%%%%%%%%%%%%%%%%%%%%%%%%%%%%
\subsection{Higgs-lepton inflation}
%%%%%%%%%%%%%%%%%%%%%%%%%%%%%%%%%%%%%%%%%%%%

The HLI model assumes that slow roll takes place along the up-type Higgs doublet-lepton doublet 
($L$-$H_{u}$) D-flat direction of the supersymmetric seesaw model.
Parametrizing this direction using a superfield $\varphi$ as
\begin{align}
L = \frac{1}{\sqrt{2}}\left(
\begin{array}{c}
\varphi \\
0
\end{array}\right), \qquad
H_{u} = \frac{1}{\sqrt{2}}\left(
\begin{array}{c}
0 \\
\varphi
\end{array}\right),
\end{align}
the superpotential becomes, ignoring $Q,~u^c,~d^c,~e^c,$ and $H_{d}$ that do not play any role during inflation,
\begin{align}
W = \frac{1}{2}MN_{R}^{c}N_{R}^{c}+\frac{1}{2}y_{D}N_{R}^{c}\varphi^{2}\,.
\end{align}
This is embedded in supergravity with the K\"{a}hler potential (in the superconformal framework)
$K = -3\Phi$, where the real function $\Phi$ is chosen to be
\begin{align}
\Phi = 1-\frac{1}{3}\left(
|N_{R}^{c}|^{2}+|\varphi|^{2}\right)
+\frac{1}{4}\gamma\left(
\varphi^{2}+\rm{c.c.}
\right)
+\frac{1}{3}\upsilon|N_{R}^{c}|^{4}\,,
\label{eqn:PhiHLI}
\end{align}
with $\gamma, \upsilon\in {\B R}$.
The term proportional to $\gamma$ violates the R parity (which is benign \cite{Arai:2012em}), and the one proportional
to $\upsilon$ represents a higher-dimensional term that controls the inflaton trajectory.
For simplicity, we consider only one generation of the right-handed neutrino\footnote{
It is straightforward to extend this model to the phenomenologically realistic cases of two or three generations of the right-handed neutrinos \cite{Arai:2012em}.
} and take $y_{D}$ to be real.

Introducing real scalar fields $s$ and $h$ by $\varphi = \frac{1}{\sqrt{2}}h$ and 
$N_{R}^{c}=\frac{1}{\sqrt{2}}s$ (here $\varphi$ and $N_{R}^{c}$ are understood as the scalar components), 
the scalar-gravity part of the Lagrangian reads
\cite{superconformal}
%\cite{Kaku:1978nz, Siegel:1978mj, Cremmer:1982en, Ferrara:1983dh, Kugo:1982mr, Kugo:1982cu, Kugo:1983mv}
%
\begin{align}
\mathcal{L}_{\rm{J}} =\sqrt{-g_{\rm{J}}}\left[
\frac{1}{2}\Phi {\C R}_{\rm{J}} - \frac{1}{2}g_{\rm{J}}^{\mu\nu}\partial_{\mu}h\partial_{\nu}h-\frac{1}{2}\kappa g_{\rm{J}}^{\mu\nu}\partial_{\mu}s\partial_{\nu}s-V_{\rm{J}}
\right],\label{eqn:LagJ}
\end{align}
where
\begin{align}
\kappa = 1-2\upsilon s^{2},
\qquad 
\xi \equiv \frac{1}{4}\gamma-\frac{1}{6}\,,
\label{eqn:kappaandxi}
\end{align}
and $\Phi$ is given by Eq. \eqref{eqn:Phi}.
For $\upsilon\neq 0$, the K\"{a}hler metric is nontrivial.
The potential is found to be
\begin{align}
V_{\rm{J}} &= \frac{1}{4}y_{D}^{2}s^{2}h^{2}
+\frac{(2\sqrt{2}Ms+y_{D}h^{2})^{2}}{16(1-2\upsilon s^{2})}
\nonumber\\
&
-\frac{1}{8}\frac{s^{2}\left(\sqrt{2}Ms+3\gamma y_{D}h^{2}-\frac{\upsilon s^{2}(y_{D}h^{2}+2\sqrt{2}Ms)}{1-2\upsilon s^{2}}\right)^{2}}{12+\frac{\upsilon s^{4}}{1-2\upsilon s^{2}}+3\gamma h^{2}(\frac{3}{2}\gamma-1)}\,.\label{eqn:VJ}
\end{align}
The Lagrangian \eqref{eqn:LagJ} involves nonminimal coupling of the scalar fields to gravity
(the subscript J stands for the Jordan frame).
Upon Weyl rescaling of the metric, one may go to the Einstein frame in which the scalars are minimally
coupled to gravity.
The resulting Lagrangian is the one we saw at the beginning \eqref{eqn:Lag}, with the scalar potential in the Einstein frame given by $V(\phi^I)=\Phi^{-2}V_{\rm{J}}$.

One can see from Eqs. \eqref{eqn:Phi} and \eqref{eqn:VJ} that the shape of the potential $V(\phi^I)$ is controlled by the four parameters $M$, $y_D$, $\gamma$ (or $\xi$), and $\upsilon$, among which $y_D$ is determined by the seesaw relation \eqref{eqn:seesaw} from $M$.
Moreover, the amplitude of the curvature perturbation (we use the Planck normalization
$A_s=2.215\times 10^{-9}$ \cite{Ade:2013uln}) provides constraints on the shape of the potential; we use this condition to fix the value of $\xi$ for a given number of $e$-folds $N_e$ 
(see Sec. \ref{sec:singlefield} below).
Thus, when $N_e$ and $M$ are given, the potential depends only on $\upsilon$.
In Fig.~\ref{fig:PotView}, we depict the shape of the potential in the Einstein frame
$V(\phi^I)$ when $M=1$ TeV, $N_e=60$, and $\upsilon$ is varied as $\upsilon=0$, $0.055$, $0.1$.
For large values of $\upsilon$, the $s$ field becomes massive and the inflaton
trajectory is forced to lie along the $s=0$ direction.
This feature is used in the previous studies of the supersymmetric Higgs inflation-type scenarios
\cite{Einhorn:2009bh,Ferrara:2010yw,Ferrara:2010in,
Einhorn:2012ih,Pallis:2011gr,Arai:2011nq,Arai:2011aa,Arai:2012em,Kawai:2014doa}
where only single-field inflation was considered.
For smaller $\upsilon$, an inflaton trajectory is curved (see Fig. \ref{fig:PotView}), and the single-field inflation picture breaks down.
While it is possible to consider the single-field case by introducing a large enough quartic K\"{a}hler term, it would certainly be important to investigate what will happen to the cosmological observables when $\upsilon$ is smaller and the multifield effects are not negligible.
Before starting to study the multifield case in Sec.~\ref{sec:deltaN}, we shall
briefly review the prediction of this model in the single-field limit.

%%%%%%%%%%%%%%%%%%%%%%%%%%%%%%%%%%%%%%%%%%%%
\subsection{Prediction in single-field limit}\label{sec:singlefield}
%%%%%%%%%%%%%%%%%%%%%%%%%%%%%%%%%%%%%%%%%%%%

%%%%%%%%%%%%%%%%%%%%%%%%%%%%%%%%%%%%%%%%%%
\begin{figure}[t]
\includegraphics[width=89mm]{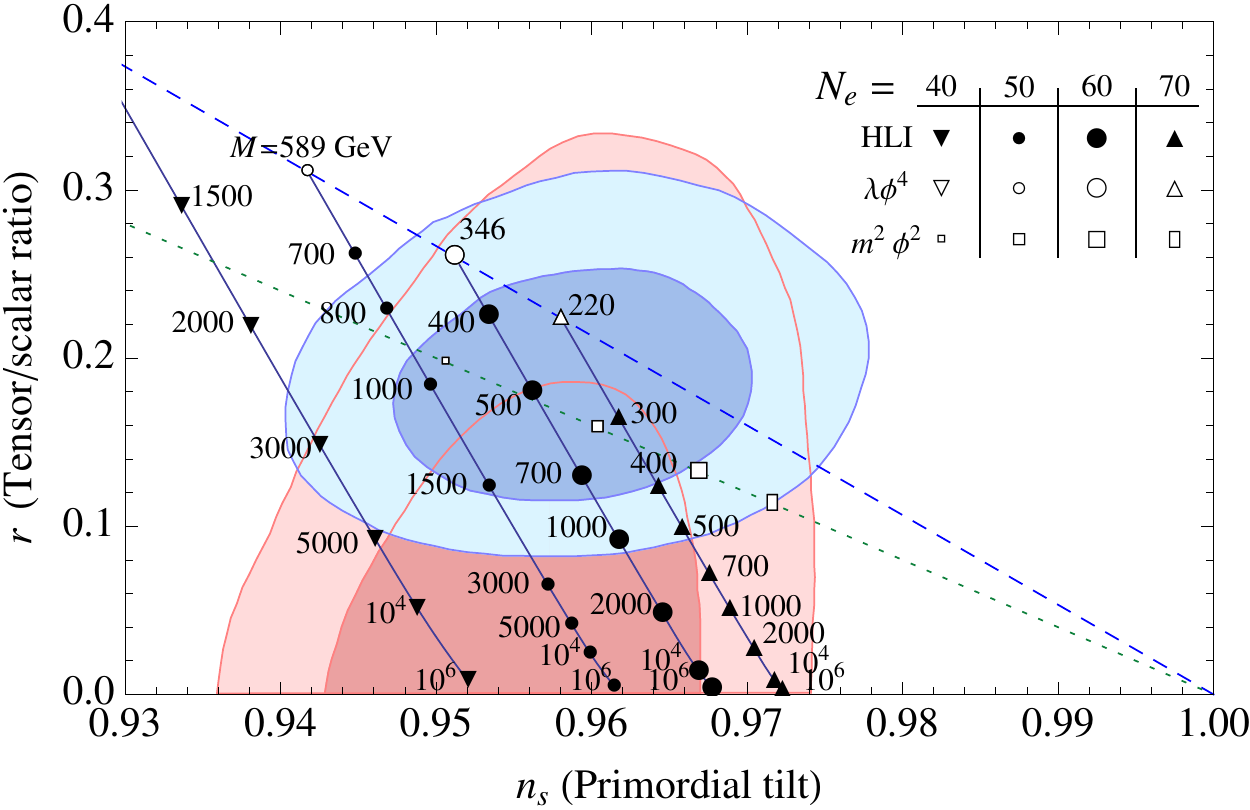}
%\includegraphics[width=89mm]{HLI_BICEP2Planck.eps}
% Here is how to import EPS art
\caption{The scalar spectral index $n_s$ and the tensor-to-scalar ratio $r$ computed in the
single-field (straight trajectory) limit. The prediction of the HLI model is the dark blue lines, for 
$e$-folding number $N_e=40, 50, 60, 70$ from the left.
The numbers shown alongside are the seesaw scale $M$ %(the mass of the right-handed neutrino)
in GeV. 
The parameter $\xi$ is fixed by the amplitude of the density fluctuations.
The contours on the background are the 68\% and the 95\% C.L. from the Planck
(Planck+WP+highL) \cite{Ade:2013zuv} (shown in red) and the BICEP2 (Planck+WP+highL+BICEP2) 
\cite{Ade:2014xna} experiments (blue).
The blue dashed line indicates the prediction of the minimally coupled $\lambda\phi^4$ model
[$n_s=1-3/(N_e+\frac 32)$, $r=\frac{16}{3}(1-n_s)$], and the green dotted line is the minimally coupled
$m^2\phi^2$ model [$n_s=1-2/(N_e+\half), r=4(1-n_s)$].
\label{fig:HLInsr_BICEP2}
}
\end{figure}
%%%%%%%%%%%%%%%%%%%%%%%%%%%%%%%%%%%%%%%%%%

%%%%%%%%%%%%%%%%%%%%%%%%%%%%%%%%%%%%%%%%%%%%
%%%%%%%%%%%%%%%%%%%%%%%%%%%%%%%%%%%%%%%%%%%%

\begin{table}[b]
\centering
\begin{tabular}{c | c | c | c | c}
\hline \hline 
\multirow{2}{*}{$M$ (GeV)} & \multirow{2}{*}{$y_{D}$} &\multicolumn{3}{c}{$\xi$}\\
\cline{3-5}
  & &  $N_{\rm{e}}=50$&  $N_{\rm{e}}=60$& $N_{\rm{e}}=70$  
\\[0.5ex] \hline \hline 
&&&&\\
$10^{3}$ & $1.277 \times 10^{-6}$ &~ $0.001588$ ~&~ $0.003696$ ~&~ $0.005957$\\[1ex]
$10^{4}$ & $4.039 \times 10^{-6}$& $0.03333$ & $0.04612$ & $0.05956$\\[1ex]
$10^{6}$ & $4.039\times 10^{-5}$& $0.7862$ & $0.9490$ & $1.112$\\[1ex]
$10^{12}$ & $0.04039$ & $868.4$ & $1031$ & $1194$\\[1ex]
\hline
\end{tabular}
\caption{
The seesaw mass $M$, the Dirac Yukawa coupling $y_D$ and the nonminimal coupling parameter $\xi$.
These quantities are related one-to-one due to the seesaw relation \eqref{eqn:seesaw} and the
Planck normalization of the curvature perturbation.
\label{table:MyDxi}
}
\end{table}
%

%%%%%%%%%%%%%%%%%%%%%%%%%%%%%%%%%%%%%%%%%%%%
%%%%%%%%%%%%%%%%%%%%%%%%%%%%%%%%%%%%%%%%%%%%

This type of inflationary model has been analyzed in detail in the single-field limit
\cite{Ferrara:2010yw,Ferrara:2010in,Pallis:2011gr,Arai:2011nq,Arai:2011aa,Arai:2012em,Kawai:2014doa} (see also Refs. \cite{Einhorn:2009bh,Einhorn:2012ih}).
Here, we summarize the prediction of the HLI model.
When the $s$ field is stablized at $s=0$, the potential \eqref{eqn:VJ} dramatically simplifies,
%\begin{align}
%V(\phi^I)\to\frac{y_D^2h^4}{16 (1+\xi h^2)^2}.
%\end{align}
and the Lagrangian \eqref{eqn:Lag} becomes
\begin{align}
{\C L}=\sqrt{-g}\Big[\half {\C R}-\frac{1+(1+6\xi)\xi h^2}{2(1+\xi h^2)^2} 
g^{\mu\nu}\partial_\mu h\partial_\nu h\cr
-\frac{y_D^2h^4}{16 (1+\xi h^2)^2}\Big].
\label{eqn:1fieldLag}
\end{align}
This is the Lagrangian of the nonminimally coupled $\lambda\phi^4$ model \cite{Okada:2010jf},
which includes the nonminimally coupled Standard Model Higgs inflation model 
\cite{CervantesCota:1995tz,Bezrukov:2007ep} as a special case.
Note that the Einstein frame Lagrangian \eqref{eqn:1fieldLag} can be obtained directly by 
Weyl transforming the Jordan frame Lagrangian
\begin{align}
{\C L}_{\rm J}=\sqrt{-g_{\rm J}}\Big[\frac{1+\xi h^2}{2}{\C R}_{\rm J} -\half g_{\rm J}^{\mu\nu}\partial_\mu h\partial_\nu h-\lambda h^4\Big],
\end{align}
where we identify $\lambda\equiv\frac{y_D^2}{16}$.

The single-field model \eqref{eqn:1fieldLag} contains two real parameters $\xi$ and $y_D$.
These are not independent, given that the amplitude of the primordial curvature perturbation is normalized as $A_s=2.215\times 10^{-9}$ \cite{Ade:2013uln}.
The model is then parametrized by (say) $y_D$ only, for a given number of $e$-folds $N_e$.
The value of $\lambda=\frac{y_D^2}{16}$ depends on phenomenological setup underlying the inflationary model.
In the case of Standard Model Higgs inflation, the parameter $\lambda$ is the Higgs self-coupling,
$\lambda\sim {\C O}(1)$ in the low energies, which gives $\xi\sim 10^2-10^4$ (with the renormalization group effects taken into account).
In the HLI model, in contrast, there are no severe experimental constraints on the Dirac Yukawa coupling $y_D$, and correspondingly the value of the nonminimal coupling parameter can be $\xi \lesssim {\C O}(1)$.
In Table \ref{table:MyDxi}, we list the values of $M$, $y_D$, and $\xi$ for $N_e=50$, 60, and 70
$e$-foldings.

Once the shape of the single-inflaton potential in the Einstein frame is determined, the slow-roll paradigm gives
a prediction for the spectra of the primordial fluctuations.
The values of the scalar spectral index $n_s$ and the tensor-to-scalar ratio $r$ in the single-field
case are shown in Fig.~\ref{fig:HLInsr_BICEP2}, for given values of the $e$-folding number $N_e$ and the seesaw mass parameter $M$ [recall that $M$ and $y_D$ are related by Eq. \eqref{eqn:seesaw}].
The uppermost points ($M=589$ GeV, 346 GeV, and 220 GeV) correspond to minimal coupling, $\xi=0$. 
Also shown on the background are the 68\% and 95\% C.L. contours from the Planck
(Planck+WP+highL, red) \cite{Ade:2013zuv} and the BICEP2 (Planck+WP+highL+BICEP2, blue) 
\cite{Ade:2014xna} experiment.
An interesting feature of the HLI model is that observation of the tensor-to-scalar ratio gives the seesaw scale $M$.
Small values of $r$ (the observation by the Planck satellite) indicate a large seesaw scale, whereas
large $r$ implied by the BICEP2 experiments suggests small seesaw scale.
It has been also pointed out that, if the underlying theory is the type-III seesaw mechanism, the parameter region
favored by BICEP2 falls into an interesting mass range that can be searched by the LHC at the 14 TeV run \cite{Kawai:2014doa}.
The HLI model also has other salient features.
It is based on the well-motivated supersymmetric seesaw model of particle physics, 
naturally explaining the small neutrino masses through the seesaw mechanism~
%\cite{Minkowski:1977sc, Yanagida:1979as, GellMann:1980vs, Mohapatra:1979ia};
\cite{seesaw};
the unitarity problem of the Standard Model Higgs inflation 
\cite{Barbon:2009ya, Burgess:2010zq, Burgess:2009ea, Hertzberg:2010dc, Bezrukov:2010jz, Lerner:2010mq,Ren:2014sya} 
is alleviated as the coupling $\xi$ can be small;
%\cite{Bezrukov:2007ep, Barvinsky:2008ia, DeSimone:2008ei, Barvinsky:2009fy}); 
leptogenesis can be implemented.
The model can be tested by the future ground-based, satellite and collider experiments, and thus
the prediction of the model deserves careful study.
In the next section, we explain the $\delta N$ formalism on which our numerical study of 
the two-field inflation model is based.

%%%%%%%%%%%%%%%%%%%%%%%%%%%%%%%%%%%%%%%%%%%%
%%%%%%%%%%%%%%%%%%%%%%%%%%%%%%%%%%%%%%%%%%%%
\section{$\delta N$ formalism}\label{sec:deltaN}
In this section, we collect elements of cosmological perturbation theory that are needed for our numerical study.
Our computation of various cosmological observables is based on the $\delta N$ formalism
%$\delta N$-formalism
\cite{Sasaki:1995aw,Nakamura:1996da,Sasaki:1998ug,Lyth:2004gb,Lyth:2005fi,Seery:2005gb,
% Backward
Yokoyama:2007dw, Yokoyama:2007uu, Yokoyama:2008by},
which has become standard for studying multifield inflation.
The $\delta N$ formalism is particularly powerful for analyzing the superhorizon 
evolution of the curvature perturbation, which is our main focus.
We essentially follow the notations of Refs. \cite{Yokoyama:2007dw, Yokoyama:2007uu} (see also Refs.
\cite{Yokoyama:2008by, Gao:2014fva}).

%%%%%%%%%%%%%%%%%%%%%%%%%%%%%%%%%%%%%%%%%%%%
%%%%%%%%%%%%%%%%%%%%%%%%%%%%%%%%%%%%%%%%%%%%

%%%%%%%%%%%%%%%%%%%%%%%%%%%%%%%%%%%%%%%%%%%%
%%%%%%%%%%%%%%%%%%%%%%%%%%%%%%%%%%%%%%%%%%%%
\subsection{Backward formalism}
We shall start by writing the background Klein--Gordon equation and the Friedmann equation in the following forms,
\begin{gather}
\frac{d^{2}\phi^{I}}{dN^{2}}+\Gamma^{I}_{JK}\frac{d\phi^{J}}{dN}\frac{d\phi^{K}}{dN}
+\left(3+\frac{1}{H}\frac{dH}{dN}\right)\frac{d\phi^{I}}{dN}
+\frac{G^{IJ}\partial_{J}V}{H^{2}} = 0\,, \\
H^{2}=\frac{1}{3}\left(
V+\frac{1}{2}H^{2}G_{IJ}\frac{d\phi^{I}}{dN}\frac{d\phi^{J}}{dN}
\right)\,,
\end{gather}
where $I=1,2$.
We have chosen the $e$-folding number $N$ defined by $dN=Hdt$ as the time variable.
The $e$-folding number $N_e$ in the previous section is $N_e=N|_{\text{end of inflation}}-N_*$,
where $N_*$ is the $e$-folding number at the horizon exit of the CMB scale.
The metric $G_{IJ}$ and the Christoffel symbol $\Gamma^I_{JK}$ of the field space are given by Eqs.
\eqref{eqn:FSmetric} and \eqref{eqn:Gamma}.
Renaming the field value and the field velocity as
\begin{align}
\varphi_{1}^{I}\equiv\phi^{I}\,,\qquad
\varphi_{2}^{I}\equiv \frac{d\phi^{I}}{dN}=\frac{d\varphi_{1}^{I}}{dN}\,,
\end{align}
the background equations of motion become
\begin{align}\label{eqn:dyneqn}
F_{1}^{I} &\equiv \frac{d\varphi_{1}^{I}}{dN}=\varphi_{2}^{I}\,,\nonumber\\
F_{2}^{I} &\equiv \frac{D\varphi_{2}^{I}}{dN} = 
-\frac{V}{H^{2}}\varphi_{2}^{I}-\frac{V^{I}}{H^{2}}\,,
\end{align}
and
\begin{align}\label{eqn:Hubble}
H^{2} &= \frac{2V}{6-G_{IJ}\varphi_{2}^{I}\varphi_{2}^{J}}\,.
\end{align}
Here, $V^{I}\equiv G^{IJ}\partial_{J}V$, and $D$ denotes the covariant derivative in the
field space, i.e.,
$D\varphi_{2}^{I} = d\varphi_{2}^{I}+\Gamma^{I}_{JK}\varphi_{2}^{J}d\varphi_{1}^{K}$.
Following Refs. \cite{Yokoyama:2007dw, Yokoyama:2007uu, Yokoyama:2008by}, we introduce a compact notation,
\begin{align}
X^{a} \equiv X^{I}_{i}\,.
\end{align}
The slow-roll parameters are defined as\footnote{
These slow-roll parameters are convenient for multifield inflation.
In the single-field limit, they are related to
$\epsilon_V\equiv M_{\rm P}^2(V'/V)^2/2$,
$\eta_V\equiv M_{\rm P}^2 V''/V$ by
$\epsilon=\epsilon_V$, and $\eta=4\epsilon_V-2\eta_V$.
In Fig. \ref{fig:HLInsr_BICEP2}, we used $\max\{\epsilon_V,|\eta_V|\}=1$ as the condition for the end 
of slow roll.
}
\begin{align}
\epsilon \equiv -\frac{\dot{H}}{H^{2}}=-\frac{1}{H}\frac{dH}{dN}\,,\qquad
\eta \equiv \frac{\dot{\epsilon}}{\epsilon H}\,
\end{align}
(the dot denotes a derivative with respect to the cosmic time),
and we assume $\epsilon,~|\eta| < 1$ during inflation and $\mathrm{max}\{\epsilon,|\eta|\}=1$ at the end of inflation. 
Temporal violation of the slow-roll approximation is possible as long as the number of $e$-foldings is not affected significantly, but we do not need to consider such cases in our study.

Based on the $\delta N$ formalism, the curvature perturbation $\zeta$ is related to the difference of the number of $e$-foldings between an initial flat hypersurface and a final uniform energy density hypersurface. We take the initial flat hypersurface to be at the Hubble exit time (i.e., $N=N_{*}$) and the final time to be $N=N_{c}$ after which the background trajectories converge. The curvature perturbation, then, will remain constant for $N>N_{c}$. Thus, we find
\begin{align}\label{eqn:curvature1}
\zeta(N_{c}) &\approx \delta N(N_{c},\varphi(N_{*})) \nonumber\\
&=N_{a}^{*}\delta\varphi_{*}^{a}+\frac{1}{2}N_{ab}^{*}\delta\varphi_{*}^{a}\delta\varphi_{*}^{b}+\cdots\,,
\end{align}
where $\delta\varphi_{*}^{a}\equiv\delta\varphi^{a}(N_{*})$ is the perturbation evaluated at the initial flat hypersurface, $N_a\equiv D N/\partial\varphi^a$, etc.

The perturbations of the scalar fields on the constant energy density hypersurface, i.e., in the
$N=$ constant gauge, are given by
\begin{align}
\delta\varphi^{a}(\lambda,N) \equiv \varphi^{a}(\lambda+\delta\lambda,N)-\varphi^{a}(\lambda,N)\,,
\end{align}
where $\lambda$'s are the $2n-1$ integration constants for an $n$-component scalar field
(we consider $n=2$), parametrizing the initial values of the fields \cite{Yokoyama:2007dw, Yokoyama:2007uu}.
We will be interested in cosmological observables up to the bispectrum of the curvature perturbation.
For our purposes, it is convenient to decompose $\delta\varphi^{a}$ into the first- and second-order quantities as
\begin{align}
\delta\varphi^{a}=\overset{(1)}{\delta\varphi^{a}}+\frac{1}{2}\overset{(2)}{\delta\varphi^{a}}\,.
\end{align}
Perturbing the set of the background equations of motion 
%ordinary differential equations 
\eqref{eqn:dyneqn},
%for a non-flat field-space metric, 
we obtain
\begin{align}\label{eqn:1stevoleqn}
\frac{D}{dN}\overset{(1)}{\delta\varphi^{a}}(N)
&=P^{a}{}_{b}(N)\,\overset{(1)}{\delta\varphi^{b}}(N)\,,
\end{align}
and
\begin{align}\label{eqn:2ndevoleqn}
\frac{D}{dN}\overset{(2)}{\delta\varphi^{a}}(N)
&=P^{a}{}_{b}(N)\,\overset{(2)}{\delta\varphi^{b}}(N)
+Q^{a}{}_{bc}(N)\,\overset{(1)}{\delta\varphi^{b}}(N)\overset{(1)}{\delta\varphi^{c}}(N)\,.
\end{align}
The explicit forms of $P^{a}{}_{b}$ and $Q^{a}{}_{bc}$ are given in Appendix \ref{apdx:expressions}.

We may write down formal solutions of Eqs. \eqref{eqn:1stevoleqn} and \eqref{eqn:2ndevoleqn} as
\begin{align}
\overset{(1)}{\delta\varphi^{a}}(N)&=\Lambda^{a}{}_{b}(N,N_{*})\overset{(1)}{\delta\varphi^{b}}(N_{*})\,,\nonumber\\
\overset{(2)}{\delta\varphi^{a}}(N)&=\int_{N_{*}}^{N}\,dN^{\prime}\,
\Lambda^{a}{}_{b}(N,N^{\prime})Q^{b}{}_{cd}(N^{\prime})\overset{(1)}{\delta\varphi^{c}}(N^{\prime})\overset{(1)}{\delta\varphi^{d}}(N^{\prime})\,,
\end{align}
where $\Lambda^{a}{}_{b}$ satisfies
\begin{align}
\frac{D}{dN}\Lambda^{a}{}_{b}(N,N^{\prime})
=P^{a}{}_{c}(N)\Lambda^{c}{}_{b}(N,N^{\prime})\,
\end{align}
and $\Lambda^{a}{}_{b}(N,N)=\delta^{a}_{b}$. Here, we have chosen $\lambda^{a}=\varphi^{a}(N_{*})$ so that we have $\delta\varphi^{a}(N_{*})=\delta\lambda^{a}$. Thus, the second-order perturbation vanishes at $N=N_{*}$.

Now, if we take $N_{F}$ to be some time later during the scalar dominant phase, the curvature perturbation is rewritten as follows:
\begin{align}\label{eqn:curvature2}
\zeta(N_{c}) &\approx \delta N(N_{c},\varphi(N_{F})) \nonumber\\
&=N_{a}^{F}\delta\varphi_{F}^{a}+\frac{1}{2}N_{ab}^{F}\delta\varphi_{F}^{a}\delta\varphi_{F}^{b}+\cdots\,.
\end{align}
Comparing with Eq. \eqref{eqn:curvature1}, we have
\begin{align}
N_{a}^{*} = N_{b}^{F}\,\Lambda^{b}{}_{a}(N_{F},N_{*})
\end{align}
and
\begin{align}
N_{ab}^{*}&=N_{cd}^{F}\,\Lambda^{c}{}_{a}(N_{F},N_{*})\Lambda^{d}{}_{b}(N_{F},N_{*})
\nonumber\\
+&2\int_{N_{*}}^{N_{F}}\,dN^{\prime}\,N_{c}(N^{\prime})Q^{c}{}_{de}(N^{\prime})
\Lambda^{d}{}_{a}(N^{\prime},N_{*})\Lambda^{e}{}_{b}(N^{\prime},N_{*})\,.
\end{align}
It is convenient to introduce a quantity $\Theta$, defined by
\begin{align}
\Theta^{a}(N)\equiv \Lambda^{a}{}_{c}(N,N_{*})A^{cb}N_{b}^{*}\,,
\end{align}
where $A^{ab}$ is the normalization factor of the two-point correlation function 
$\langle\delta\varphi_*^a\delta\varphi_*^b\rangle$ including the slow-roll corrections \cite{Nakamura:1996da, Byrnes:2006vq, Byrnes:2006fr}. 
The definition and the explicit forms of $A^{ab}$ are given in Appendix \ref{apdx:expressions}.
Then, $N_{a}(N)$ and $\Theta^{a}(N)$ satisfy the following equations:
\begin{align}\label{eqn:evoleqn}
\frac{D}{dN}N_{a}(N)&=-N_{b}(N)P^{b}{}_{a}(N)\,,\nonumber\\
\frac{D}{dN}\Theta^{a}(N)&=P^{a}{}_{b}(N)\Theta^{b}(N)\,.
\end{align}
Following the prescription of Ref. \cite{Yokoyama:2007dw}, we first solve the first equation of Eq. \eqref{eqn:evoleqn} backward until $N=N_{*}$, with the initial conditions $N_{a}(N_{F})=N_{a}^{F}$. Then, with the initial conditions $\Theta^{a}(N_{*})=A^{ab}N_{b}^{*}$, we solve the second equation of Eq. \eqref{eqn:evoleqn} forward until $N=N_{F}$.

The explicit expressions for $N_{a}^{F}$ and $N_{ab}^{F}$, which are presented in Appendix~\ref{apdx:inicond}, may be obtained by using the fact that the uniform energy density hypersurface is equivalent to the constant Hubble hypersurface on the superhorizon scales~\cite{Yokoyama:2007dw} (see also Refs. \cite{Anderson:2012em, Elliston:2012ab}),
\begin{align}
H(\varphi^{a}(N_{F}+\zeta(N_{F}))) = H(\overset{(0)}{\varphi^{a}}(N_{F}))\,,
\end{align}
where $\overset{(0)}{\varphi^{a}}$ are the background trajectories. Note that $N_{F}$ is a uniform energy density hypersurface, and we neglect the later evolution of the curvature perturbations~\cite{Yokoyama:2007dw}.

%%%%%%%%%%%%%%%%%%%%%%%%%%%%%%%%%%%%%%%%%%%%
%%%%%%%%%%%%%%%%%%%%%%%%%%%%%%%%%%%%%%%%%%%%

%%%%%%%%%%%%%%%%%%%%%%%%%%%%%%%%%%%%%%%%%%%%
%%%%%%%%%%%%%%%%%%%%%%%%%%%%%%%%%%%%%%%%%%%%
\subsection{Cosmological observables}
Using the backward formalism, one can compute various cosmological observables. 
Here, we give the expressions for the scalar and tensor power spectra, the scalar and tensor spectral indices, the tensor-to-scalar ratio, and the nonlinearity parameter
\cite{Sasaki:1995aw, Yokoyama:2007dw, Yokoyama:2007uu, Yokoyama:2008by,Gao:2014fva}.

%%%%%%%%%%%%%%%%%%%%%%%%%%%%%%%%%%%%%%%%%%%%
\subsubsection{Power spectra}
In momentum space, the two-point correlator of the curvature perturbation is written as
\begin{align}
\langle\zeta_{{\BS k}_1}\zeta_{{\BS k}_2}\rangle
=(2\pi)^3\delta^3\left({\BS k}_1+{\BS k}_2\right)P_\zeta(k).
\end{align}
The power spectrum of the scalar perturbation is given by
\begin{align}
\mathcal{P}_{S}=\frac{k^3}{2\pi^2}P_\zeta(k),
\end{align}
and in the $\delta N$ formalism it is expressed as \cite{Sasaki:1995aw, Byrnes:2006fr}
\begin{align}\label{eqn:scalarPS}
\mathcal{P}_{S}=\left(\frac{H_{*}}{2\pi}\right)^{2}A^{ab}_{*}N_{a}^{*}N_{b}^{*}\,.
\end{align}
Similarly, the power spectrum of the tensor perturbation is
\begin{align}
\mathcal{P}_{T} =\frac{k^3}{2\pi^2}P_h(k),
\end{align}
where $P_h(k)$ is given by the two-point correlator of the tensor perturbation
\begin{align}
\langle h_{ij}({\BS k}_1) h^{ij}({\BS k}_2)\rangle =(2\pi)^3\delta({\BS k}_1+{\BS k}_2) P_h(k).
\end{align}
In the $\delta N$ formalism,
\begin{align}\label{eqn:tensorPS}
\mathcal{P}_{T} = 8\left(\frac{H_{*}}{2\pi}\right)^{2}\big[1+2(\alpha-1)\epsilon\big]_{*}\,,
\end{align}
where $\alpha \equiv 2-\ln 2 - \gamma_{EM} \simeq 0.7296$, 
with $\gamma_{EM} \simeq 0.5772$ the Euler--Mascheroni constant.
%%%%%%%%%%%%%%%%%%%%%%%%%%%%%%%%%%%%%%%%%%%%

%%%%%%%%%%%%%%%%%%%%%%%%%%%%%%%%%%%%%%%%%%%%
\subsubsection{Spectral indices}
The spectral index for the scalar perturbation is
\begin{align}\label{eqn:ns}
n_{s}-1 = \frac{D\ln\mathcal{P}_{S}}{d\ln k} \simeq \frac{D\ln\mathcal{P}_{S}}{dN}\,,
\end{align}
where we used $d\ln k=d\ln aH\simeq d\ln a= dN$ to obtain the last expression. 
%From the scalar power spectrum \eqref{eqn:scalarPS} 
%and the evolution equations \eqref{eqn:evoleqn} we obtain
%%
%\begin{align}
%n_{s}-1 = -2\epsilon -2\frac{A^{ab}N_{a}P^{c}_{b}N_{c}}{A^{ab}N_{a}N_{b}}
%+\frac{\left(\frac{DA^{ab}}{dN}\right)N_{a}N_{b}}{A^{ab}N_{a}N_{b}}\,.
%\end{align}
%
Similarly, the tensor spectral index is
\begin{align}\label{eqn:nt}
n_{t}
&=\frac{D\ln\mathcal{P}_{T}}{d\ln k} \simeq \frac{D\ln\mathcal{P}_{T}}{dN}
%\nonumber\\
%&
=
-2\epsilon
\frac{1-(\alpha-1)\eta}{1+2(\alpha-1)\epsilon}
%\nonumber\\
%&\simeq 
%-2\epsilon+\frac{2(\alpha-1)\epsilon\eta}{1+2(\alpha-1)\epsilon}\,.
\end{align}
It is implicit that these quantities are evaluated at $N=N_{*}$.
%%%%%%%%%%%%%%%%%%%%%%%%%%%%%%%%%%%%%%%%%%%%

%%%%%%%%%%%%%%%%%%%%%%%%%%%%%%%%%%%%%%%%%%%%
\subsubsection{Tensor-to-scalar ratio}
The tensor-to-scalar ratio is defined by
\begin{align}
r \equiv \frac{\mathcal{P}_{T}}{\mathcal{P}_{S}}\,,
\end{align}
and using Eqs. \eqref{eqn:scalarPS} and \eqref{eqn:tensorPS}, we have
\begin{align}
r = 8\frac{\big[1+2(\alpha-1)\epsilon\big]_{*}}{A^{ab}_{*}N_{a}^{*}N_{b}^{*}}\,.
\end{align}
%
%%%%%%%%%%%%%%%%%%%%%%%%%%%%%%%%%%%%%%%%%%%%

%%%%%%%%%%%%%%%%%%%%%%%%%%%%%%%%%%%%%%%%%%%%
\subsubsection{Nonlinearity parameter}
The nonlinearity parameter $f_{\rm{NL}}$ is a measure of non-Gaussianities in the primordial density fluctuations, defined by the bispectrum, i.e., the three-point correlation function of the curvature
perturbation
\begin{align}
\langle\zeta_{{\BS k}_1}\zeta_{{\BS k}_2}\zeta_{{\BS k}_3}\rangle
=(2\pi)^3\delta^3\left({\BS k}_1+{\BS k}_2+{\BS k}_3\right)
B_{\zeta}({k}_1, {k}_2, {k}_3).
\end{align}
We will be focusing on the so-called local-type nonlinearity parameter, defined through the ratio
of the bispectrum and the power spectrum as
\begin{align}
B_{\zeta}({k}_1, {k}_2, {k}_3)=\frac 65 f_{NL}^{\rm local}\Big\{P_\zeta({k}_1)P_\zeta({k}_2)
+\mbox{2 perms}\Big\}.
\end{align}
The local-type non-Gaussianity is generated by nonlinear interactions after the horizon exit
\cite{Acquaviva:2002ud, Maldacena:2002vr, Tanaka:2010km, Qiu:2010dk}. 
There are other types of non-Gaussian profiles that can be generated in different mechanisms 
(see, e.g., Ref. \cite{Kawai:2014vxa}).

The local-type nonlinearity parameter $f_{\rm{NL}}=f_{\rm{NL}}^{\rm local}$ (we will omit ``local'' hereafter) is conveniently computed using the 
$\delta N$ formalism \cite{Lyth:2005fi}, and its leading contribution (the scale-independent part) is
\begin{align}\label{eqn:fNL}
f_{\rm{NL}}\simeq f_{\rm{NL}}^{(4)} = \frac{5}{6}\frac{A^{ac}_{*}A^{bd}_{*}N_{c}^{*}N_{d}^{*}N_{ab}^{*}}{\left(A^{ab}_{*}N_{a}^{*}N_{b}^{*}\right)^{2}}\,
\end{align}
(the superscript ``$(4)$'' denotes the the scale-independent part in the convention of Refs. 
\cite{Vernizzi:2006ve, Choi:2007su}).
Other (scale-dependent) parts are subleading and will be neglected.
%%%%%%%%%%%%%%%%%%%%%%%%%%%%%%%%%%%%%%%%%%%%
%%%%%%%%%%%%%%%%%%%%%%%%%%%%%%%%%%%%%%%%%%%%

%%%%%%%%%%%%%%%%%%%%%%%%%%%%%%%%%%%%%%%%%%
%\begin{center}
\begin{figure*}[t]
%\begin{eqnarray*}
%\begin{array}{ccc}
\includegraphics[width=89mm]{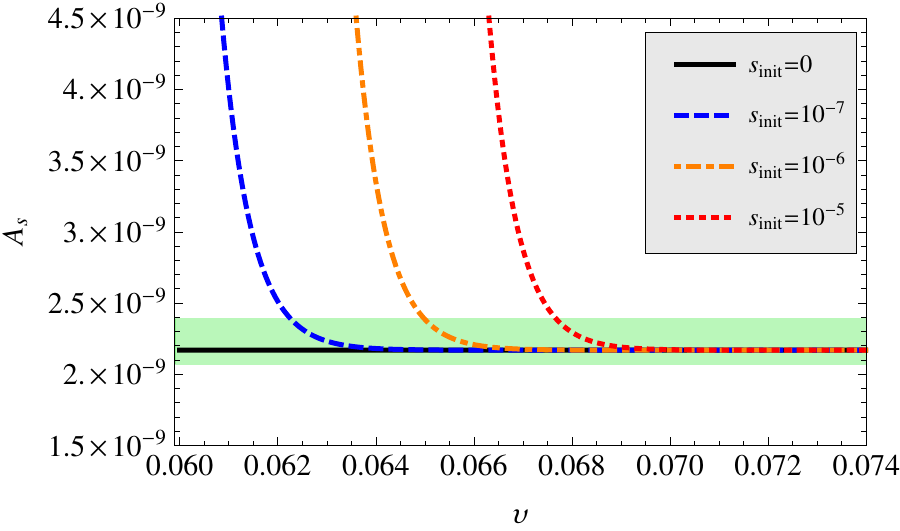}
%\includegraphics[width=89mm]{FigAs.eps}
%&&
\includegraphics[width=89mm]{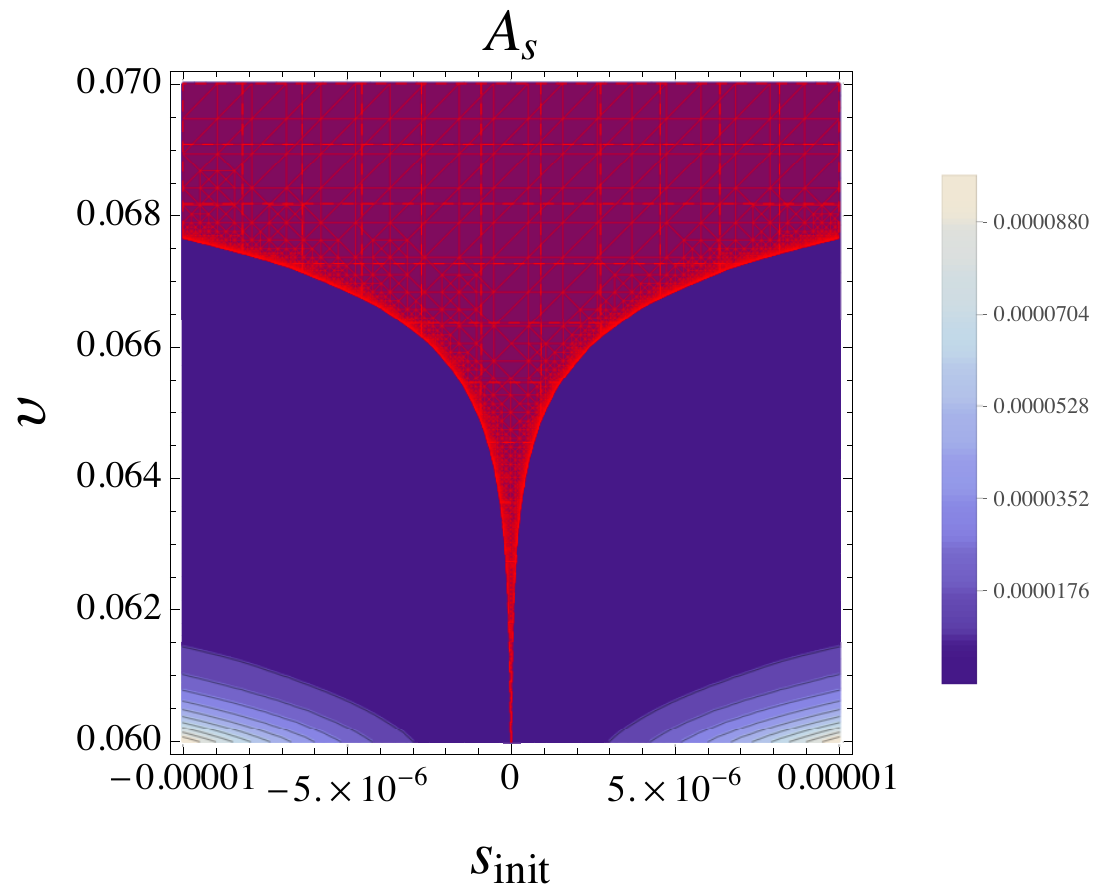}
%\includegraphics[width=89mm]{FigAscont_r.eps}
%\end{array}
%\end{eqnarray*}
\caption{\label{fig:As}
The amplitude of the scalar perturbation $A_s$ as a function of $\upsilon$ (left).
The initial condition for the $s$ field is chosen as $s_{\rm init}=0$, $1.0\times 10^{-7}$, 
$1.0\times 10^{-6}$, and $1.0\times 10^{-5}$.
The initial conditions for $\dot s_{\rm init}$, $h_{\rm init}$, and $\dot h_{\rm init}$ are the same as in Fig. \ref{fig:PotView}.
The green-shaded region is the Planck constraints \cite{Ade:2013zuv}
$A_s=(2.23\pm 0.16)\times 10^{-9}$.
The panel on the right shows the contour plot for $-10^{-5}\leq s_{\rm init}\leq 10^{-5}$ and $0.06\leq\upsilon\leq 0.07$. 
The parameter region within the Planck constraints is shaded red.
}
\end{figure*}
%\end{center}
%%%%%%%%%%%%%%%%%%%%%%%%%%%%%%%%%%%%%%%%%%

%%%%%%%%%%%%%%%%%%%%%%%%%%%%%%%%%%%%%%%%%%%%
%%%%%%%%%%%%%%%%%%%%%%%%%%%%%%%%%%%%%%%%%%%%
\section{Numerical results}\label{sec:result}
%%%%%%%%%%%%%%%%%%%%%%%%%%%%%%%%%%%%%%%%%%%%

We present the numerical results of our analysis in this section.
Before going into details, we comment on our strategy and the method.
Our analysis is based on the $\delta N$ formalism as outlined in the previous section\footnote{
We checked our numerical code on simple two-field models with separable potentials,
including Refs. \cite{GarciaBellido:1995qq,Vernizzi:2006ve,Choi:2007su,Gong:2011cd,Kim:2014vsa}.
}.
This formalism assumes complete decay of the isocurvature mode before the final hypersurface is reached.
Our results presented here also hinge upon this assumption.
The fate of the isocurvature mode is in general model dependent, and in the particular model discussed here, the dominant component of the isocurvature mode is the right-handed scalar neutrino, which is expected to decay and generate lepton numbers. 
Strictly speaking, cosmological observables in multifield scenarios depend on details of the reheating process, and hence it is important to keep in mind that our results below include uncertainty regarding this point.
See, e.g., Ref. \cite{Meyers:2013gua} for a recent study on the relation between the primordial observables 
and the decay process.

We investigate the parameter space only in the vicinity of the single-field limit of the model;
for this purpose, we choose the initial conditions for the inflaton to realize a nearly
straight trajectory.
We also use the value of the parameter $\xi$ that is fixed by the normalization of the scalar
power spectrum $A_{s}$ in the single-field limit;
this leads to slight inconsistency of the parameter choice as $A_{s}$ also changes
as other parameters ($\upsilon$ and the initial $s$-field value $s_{\rm init}$) are varied.
We shall, however, see that this is a minor issue as the change of $A_{s}$ is less significant than that of
the nonlinearity parameter $f_{\rm NL}$.
Fixing the value of $\xi$ is also convenient for observing the overall behavior of the cosmological observables without introducing complexities.

%%%%%%%%%%%%%%%%%%%%%%%%%%%%%%%%%%%%%%%%%%%%
\subsection{Generic features}

%%%%%%%%%%%%%%%%%%%%%%%%%%%%%%%%%%%%%%%%%%
%\begin{center}
\begin{figure*}[t]
%\begin{eqnarray*}
%\begin{array}{ccc}
\includegraphics[width=89mm]{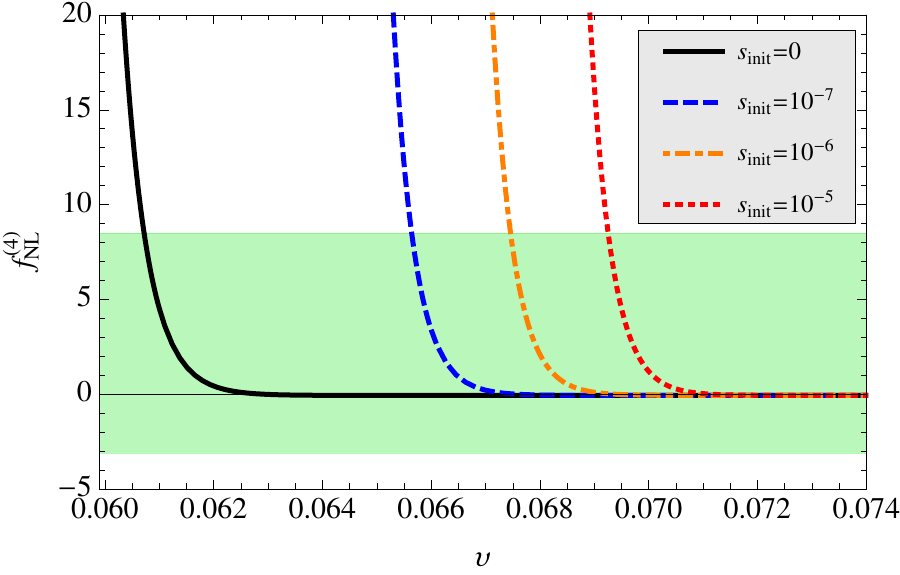}
%\includegraphics[width=89mm]{FigfNL.eps}
%&&
\includegraphics[width=89mm]{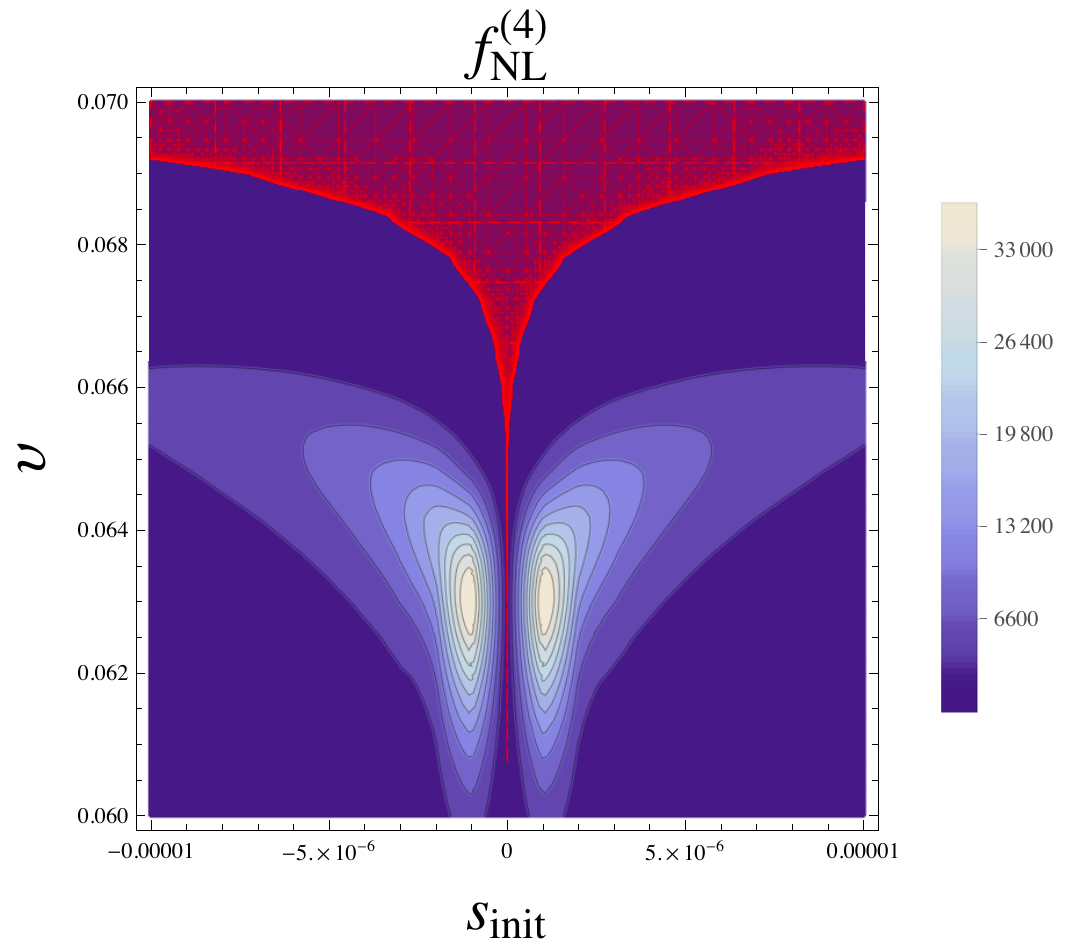}
%\includegraphics[width=89mm]{FigfNLcont_r.eps}
%&&
%\end{array}
%\end{eqnarray*}
\caption{\label{fig:fNL}
The nonlinearity parameter $f_{\rm NL}$ as a function of $\upsilon$ (left).
The initial conditions for the $s$ field are chosen as $s_{\rm init}=0$, $1.0\times 10^{-7}$, 
$1.0\times 10^{-6}$, and $1.0\times 10^{-5}$.
The initial conditions for $\dot s_{\rm init}$, $h_{\rm init}$, and $\dot h_{\rm init}$ are the same as in Fig. \ref{fig:PotView}.
The green-shaded region corresponds to the Planck constraints 
$f_{\rm NL}^{\rm local}=2.7\pm 5.8$ \cite{Ade:2013ydc}.
The right panel shows the contour plot of $f_{\rm NL}$ for $-10^{-5}\leq s_{\rm init}\leq 10^{-5}$ 
and $0.06\leq\upsilon\leq 0.07$, with the red shade indicating the parameter region allowed by
Planck.
}
\end{figure*}
%\end{center}
%%%%%%%%%%%%%%%%%%%%%%%%%%%%%%%%%%%%%%%%%%

%%%%%%%%%%%%%%%%%%%%%%%%%%%%%%%%%%%%%%%%%%%%
\subsubsection{Procedure and model parameters}

%Procedure
We compute the cosmological observables by taking the following steps.
We first fix the values of the $e$-folding number $N_e$ and the seesaw mass scale $M$.
In the single-field limit, there is no more free parameter; the end of the slow roll is 
characterized by $\max \{\epsilon, |\eta|\}=1$, and the horizon exit of the CMB scale is $N_e$ $e$-folds back in time from there. 
We denote the inflaton value at the horizon exit determined this way as $h_{\rm init}$ and
use it as the initial value for the $h$ field.
The nonminimal coupling parameter $\xi$ is fixed by the scalar power spectrum.
We next move on to the two-field model, using the same $N_e$, $M$, $h_{\rm init}$, and $\xi$ as above.
We use the initial value $\dot h_{\rm init}$ for $\dot h$ (derivative with respect to the cosmic time) given by the slow-roll equation of motion for the $h$ field (see below), and we set the initial velocity for the $s$ field to be $\dot s_{\rm init}$=0.
We shall adjust the initial value $s_{\rm init}$ of the $s$ field to see its effects on the observables.
We then solve the equations of motion forward in time (down the potential), using a certain value of the K\"{a}hler potential parameter $\upsilon$.
The resulting trajectory is generally curved; we identify the end of the slow roll using the
condition $\max\{\epsilon,|\eta|\}=1$  (for the two-field model) and denote the field values there as
$s_{\rm end}$, $h_{\rm end}$.
We then solve the equations of motion backward
in time from $s_{\rm end}$, $h_{\rm end}$ for $N_e$ $e$-folds, to find the (two-field) horizon exit values of the fields $s_*$ and $h_*$.
Finally, we follow the trajectory forward in time to find the power spectra, the nonlinearity parameter, etc.

In the numerical study below, we use the $e$-folding number $N_e=60$ and the seesaw mass scale
$M=1\,\rm{TeV}$ (we also made computation for other parameters; see the comments below).
These are in the parameter range that is interesting in both cosmology and in particle phenomenology.
For the value of $\xi$, we use $\xi= 3.696\times 10^{-3}$, which is determined by the Planck normalization of the scalar power spectrum (we use the Planck $+$ WP best-fit value 
\cite{Ade:2013zuv} $A_{s} = 2.215\times 10^{-9}$); see Table \ref{table:MyDxi}.
The Dirac Yukawa coupling $y_{D}$ is fixed by the seesaw relation \eqref{eqn:seesaw}.
The parameters $s_{\rm init}$ and $\upsilon$ are to be varied.

%%%%%%%%%%%%%%%%%%%%%%%%%%%%%%%%%%%%%%%%%%%%
\subsubsection{Initial condition dependence}\label{subsec:EIC}

As shown in Fig.~\ref{fig:PotView}, the trajectory of the inflaton is sensitive to the initial conditions of the field dynamics when the value of $\upsilon$ is small.
Since our focus is on the prediction of the model as the trajectory deviates from a straight line, we shall specify the initial values as follows.
We assume that the initial value of $s$ to be small and the initial $h$ is determined by the
single-field value, e.g., $h_{\rm init}=21.99$ for $N_e=60$.
As any light field has quantum fluctuations of the order of the Hubble parameter during inflation,
we expect small but nonzero values of initial $s$.
Assuming that the $s$ field is light, the size of the fluctuations is
\begin{align}
\langle(\Delta s_{\rm can})^2\rangle\approx\langle G_{11}(\Delta s)^2\rangle\approx \frac{H^2}{(2\pi)^2},\label{eqn:Deltas}
\end{align}
where $s_{\rm can}$ is the canonically normalized $s$ field in the Einstein frame.
The Hubble parameter at the horizon exit of the CMB scale is determined by 
amplitude of the curvature perturbation and the tensor-to-scalar ratio as
\begin{align}
H\approx\frac{\pi}{M_{\rm P}}\sqrt{\frac{r A_s}{2}}\approx 0.5\times 10^{-4} M_{\rm P},
\end{align}
where we used $r\approx 0.1$. 
Combining this with Eq. \eqref{eqn:Deltas}, we evaluate the natural initial values for the $s$ field to be
in the range
\begin{align}
-\Delta s\leq s_{\rm init}\leq\Delta s,
\end{align}
where
\begin{align}
\Delta s\simeq 
\frac{H}{2\pi}\sqrt{1+\xi h}
\approx 10^{-5} M_{\rm P},
\end{align}
in the case of $M=1$ TeV and $N_e=60$.
We thus consider $s_{\rm init}$ in the range 
$-10^{-5} M_{\rm P}\leq s_{\rm init}\leq 10^{-5}M_{\rm P}$ in the following analysis.

As mentioned, the velocity of the $s$ field is set to zero, $\dot s_{\rm init}=0$, and the velocity of the $h$ field is
determined by the slow-roll equation of motion, i.e.,
\begin{align}
3H\dot h+G^{hI}\frac{\partial V}{\partial \phi^I}\Big|_{s=s_{\rm init},h=h_{\rm init}}=0.
\end{align}

%%%%%%%%%%%%%%%%%%%%%%%%%%%%%%%%%%%%%%%%%%%%
%\subsection{Effects of $\upsilon$}
\subsubsection{K\"{a}hler parameter dependence}\label{subsec:Eupsilon}
%%%%%%%%%%%%%%%%%%%%%%%%%%%%%%%%%%%%%%%%%%%%

Our K\"{a}hler potential \eqref{eqn:PhiHLI} [see also Eq. \eqref{eqn:Phi}] includes two tuneable
parameters $\gamma$ and $\upsilon$.
The former is determined by $\xi$ through Eq. \eqref{eqn:kappaandxi}.
We vary $\upsilon$ and investigate how the observables change.
Obviously, one can see from Eq. \eqref{eqn:Phi} that there are no effects of $\upsilon$ when $s=0$;
in this case, the the model becomes the nonminimally coupled $\lambda\phi^4$ model that we illustrated 
in Sec.~\ref{sec:singlefield}.
The effects of $\upsilon$ become important when the inflaton trajectory deviates from $s=0$.

As shown on the left panel of Fig.~\ref{fig:PotView}, for very small values of $\upsilon$, the 
initial value for $s$ needs to be fine-tuned to some nonzero value in order for the inflaton trajectory to reach the supersymmetric vacuum $(s,h)=(0,0)$ 
[we see in the expression \eqref{eqn:VJ} that the potential $V(\phi^I)$ is not symmetric in $s$; 
thus, a trajectory with the initial conditions $s_{\rm init}=0$ and $\dot s_{\rm inti}=0$ does not necessarily come straight down to the supersymmetric vacuum].
For larger values of $\upsilon$, the potential is stabilized in the direction of $s$, and thus the danger of the trajectory falling into an unphysical vacuum ceases to bother us.
However, a curved trajectory generally results in cosmological parameters outside the observational constraints. 
For even larger values of $\upsilon$, the inflaton trajectory becomes insensitive to the initial
conditions, and the prediction of the model converges to that of single-field inflation.
As we start from the single-field limit (large enough $\upsilon$) that agrees with observations and
tune $\upsilon$ to lower values, the prediction of the model goes outside the observational bound
at some value of $\upsilon$.
This transition takes place around $\upsilon\sim 0.0607$, for the $M=1$ TeV and $N_e=60$ case that
we consider.
While there may be islands in the parameter space that are compatible with observations, the analysis as prescribed above gives reasonable constraints on the K\"{a}hler potential in the vicinity of the straight trajectory background solutions.

%%%%%%%%%%%%%%%%%%%%%%%%%%%%%%%%%%%%%%%%%%
%\begin{center}
\begin{figure}[h]
%\begin{eqnarray*}
%\begin{array}{ccc}
\includegraphics[width=87mm]{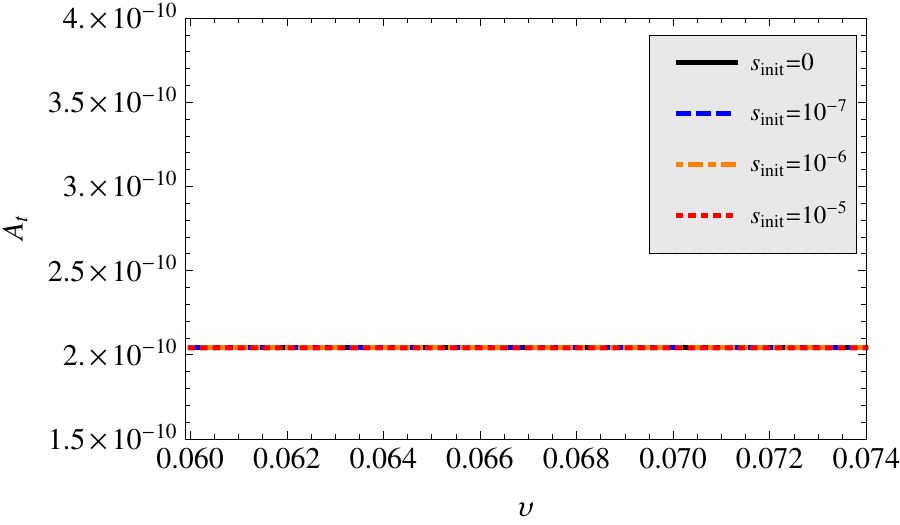}
%\includegraphics[width=87mm]{FigAt.eps}
%\end{array}
%\end{eqnarray*}
\caption{\label{fig:At}
The amplitude of the tensor perturbation $A_t$ as a function of $\upsilon$, for the initial 
conditions $s_{\rm init}=0$, $1.0\times 10^{-7}$, $1.0\times 10^{-6}$, and $1.0\times 10^{-5}$.
The initial conditions for $\dot s_{\rm init}$, $h_{\rm init}$, and $\dot h_{\rm init}$ are the same as in Fig. \ref{fig:PotView}.
The tensor mode does not interact outside the horizon and hence is insensitive to the change of the background trajectory. 
}
\end{figure}
%\end{center}
%%%%%%%%%%%%%%%%%%%%%%%%%%%%%%%%%%%%%%%%%%

%%%%%%%%%%%%%%%%%%%%%%%%%%%%%%%%%%%%%%%%%%%%
\subsection{Numerical results for cosmological parameters}
In this subsection, we describe the behavior of cosmological parameters as the values of $s_{\rm init}$ and 
$\upsilon$ are varied.

\subsubsection{Scalar power spectrum}

The scalar power spectrum \eqref{eqn:scalarPS} may be written as 
\begin{align}
{\mathcal P}_S=A_s \left(\frac{k}{k_0}\right)^{n_s-1+\half\frac{d n_s}{d\ln k}\ln\frac{k}{k_0}+\cdots},
\end{align}
where $A_s$ is the normalized amplitude at the pivot scale $k=k_0$
and $n_s$ is the scalar spectral index that will be discussed later.
This $A_s$ is to be compared with the observational constraints \cite{Ade:2013zuv}
\begin{align}\label{eqn:PlanckAs}
A_s\times 10^{9}&=2.23\pm 0.16 &\text{(Planck)},\cr 
&=2.196^{+0.051}_{-0.060} &\text{(Planck + WP)},
\end{align}
at $k_0=0.05\,\text{Mpc}^{-1}$.
In Fig.~\ref{fig:As}, we show our numerical results for the scalar power spectrum \eqref{eqn:scalarPS}. 
The panel on the left shows the values of ${\C P}_S\approx A_s$ for different initial conditions
$s_{\rm init}=0$, $1.0\times 10^{-7}$, $1.0\times 10^{-6}$, $1.0\times 10^{-5}$ and for the K\"{a}hler potential parameter 
$ 0.06\leq\upsilon\leq 0.074$.
We have chosen $M=1\,\rm{TeV}$ and $N_e=60$. 
The green-shaded region indicates the Planck constrains of Eq. \eqref{eqn:PlanckAs}.
The right panel shows a contour plot in the $s_{\rm init}$--$\upsilon$ plane. The red-shaded color indicates the allowed parameter region within the Planck constraints \eqref{eqn:PlanckAs}.

We see that, as the parameter $\upsilon$ is tuned to a smaller value, the predicted value of $A_s$ will become larger
and go out of the observational bounds.
For larger $|s_{\rm init}|$, the constraints on $\upsilon$ becomes tighter (the lower bound for $\upsilon$ becomes larger).
This can be understood as an effect of the isocurvature mode: 
the curvature perturbation at superhorizon scales is sourced by the isocurvature mode.
The conversion of power from the isocurvature mode to the curvature mode takes place when the 
trajectory is curved.
As a consequence, the curvature perturbation becomes larger at the end of inflation than at the horizon exit, and this enhancement is more efficient if the inflaton makes a sharp turn (i.e., for larger $|s_{\rm init}|$).
Because of the quantum fluctuations, uncertainty of $\Delta s_{\rm init}\sim 10^{-5}$ is expected. 
This means that fine-tuning of the initial condition for $s_{\rm init}$ to be less than $10^{-5}$ is unnatural.
We thus conclude that the constraints $A_s=(2.23\pm 0.06)\times 10^{-9}$ (Planck) give $\upsilon \gtrsim 0.06767$.
The Planck $+$ WP constraints $A_s=2.196({}^{+0.051}_{-0.060})\times 10^{-9}$ give a tighter bound, $\upsilon \gtrsim 0.06827$.

%%%%%%%%%%%%%%%%%%%%%%%%%%%%%%%%%%%%%%%%%%
%\begin{center}
\begin{figure*}[t]
%\begin{eqnarray*}
%\begin{array}{ccc}
\includegraphics[width=89mm]{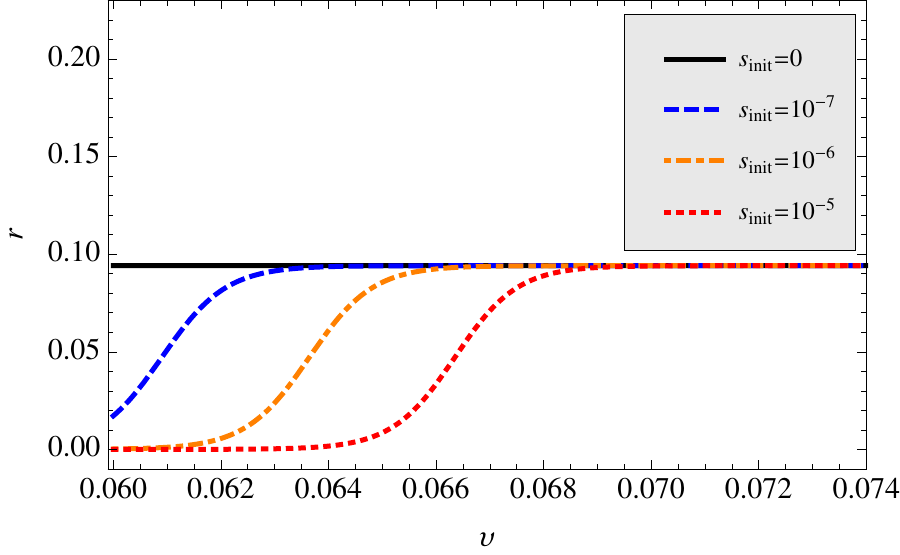}
%\includegraphics[width=89mm]{Figr.eps}
%&&
\includegraphics[width=89mm]{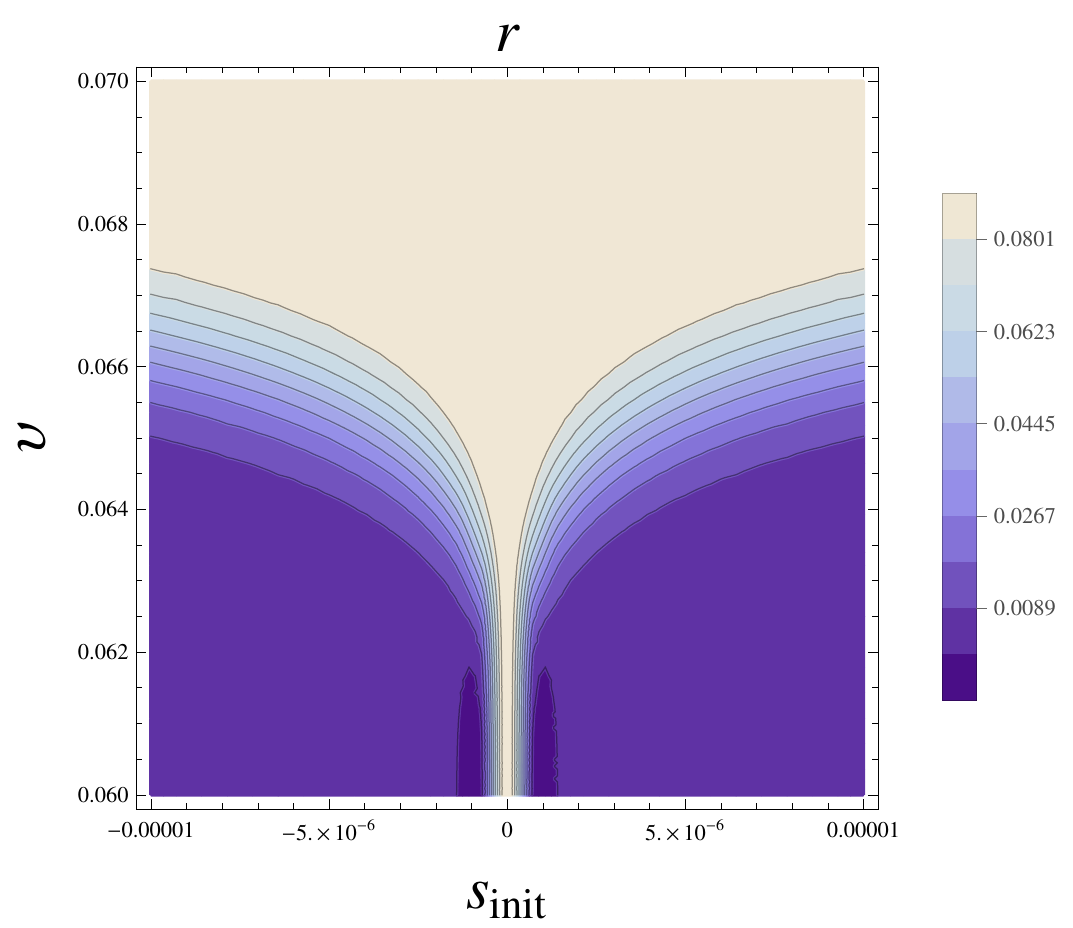}
%\includegraphics[width=89mm]{Figrcont.eps}
%&&
%\end{array}
%\end{eqnarray*}
\caption{\label{fig:r}
The tensor-to-scalar ratio $r$ as a function of $\upsilon$ for the initial values of the $s$ field
$s_{\rm init}=0$, $1.0\times 10^{-7}$, $1.0\times 10^{-6}$, and $1.0\times 10^{-5}$ (left).
The initial conditions for $\dot s_{\rm init}$, $h_{\rm init}$, and $\dot h_{\rm init}$ are the same as in Fig. \ref{fig:PotView}.
The panel on the right shows the contour plot for $r$ in the range 
$-10^{-5}\leq s_{\rm init}\leq 10^{-5}$ and $0.06\leq\upsilon\leq 0.07$.
}
\end{figure*}
%\end{center}
%%%%%%%%%%%%%%%%%%%%%%%%%%%%%%%%%%%%%%%%%%

\subsubsection{Scalar bispectrum}

Now, we turn our attention to the nonlinearity parameter $f_{\rm NL}$.
Since the main contribution comes from the scale-independent part of the local-type bispectrum, we
consider the $f_{\rm NL}^{(4)}$ given by the expression \eqref{eqn:fNL}.
The numerical results are shown in Fig. \ref{fig:fNL}.
The left panel shows $f_{\rm NL}$ for the same parameter choice as in the scalar power spectrum case above, namely, $0.06\leq\upsilon\leq 0.074$ and $s_{\rm init}=0$, $1.0\times 10^{-7}$, $1.0\times 10^{-6}$, and $1.0\times 10^{-5}$.
The nonlinearity parameter becomes large as $\upsilon$ is decreased, similarly to the $A_s$ case above.
However, $f_{\rm NL}$ is more susceptible than $A_s$ to the multifield effects and goes outside the Planck constraints at a larger value of $\upsilon$.
Both the enhancement in $A_s$ and the generation of $f_{\rm{NL}}$ are due to interaction at superhorizon scales.
However, as pointed out, e.g., in Refs. \cite{Peterson:2010mv,Kaiser:2012ak}, the primary contribution to
$f_{\rm NL}$ comes from the {\em change of} the curvature-isocurvature transfer function 
$T_{{\zeta}{\C S}}$ (proportional to $N_{ab}$), whereas the growth of $A_s$ is caused by 
$T_{{\zeta}{\C S}}$ itself ($\sim N_a$).
This observation justifies the procedure of our analysis---the value of $\xi$ was fixed by $A_s$, so when $A_s$ changes, $\xi$ needs to be readjusted; the above finding indicates that such readjustment is not necessary within the parameter range where only $f_{\rm NL}$ changes significantly.

The right panel of Fig. \ref{fig:fNL} shows a contour plot in the corresponding parameter region. 
It indicates that large non-Gaussianities are obtained within some islands of the parameter space.
This is in agreement with our understanding that the local-type non-Gaussianities are generated
at superhorizon scales by nonlinear interactions, and there is a tradeoff between generation of a sizeable isocurvature mode and efficient conversion of it into the curvature mode \cite{Peterson:2010mv,Peterson:2010np}.

%%%%%%%%%%%%%%%%%%%%%%%%%%%%%%%%%%%%%%%%%%
%\begin{center}
\begin{figure*}[t]
%\begin{eqnarray*}
%\begin{array}{ccc}
\includegraphics[width=89mm]{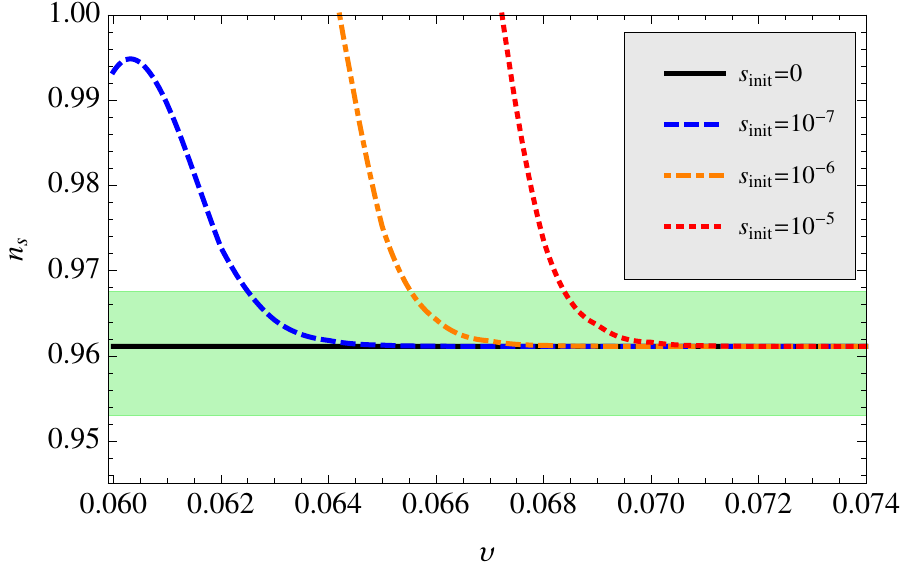}
%\includegraphics[width=89mm]{Figns.eps}
%&&
\includegraphics[width=89mm]{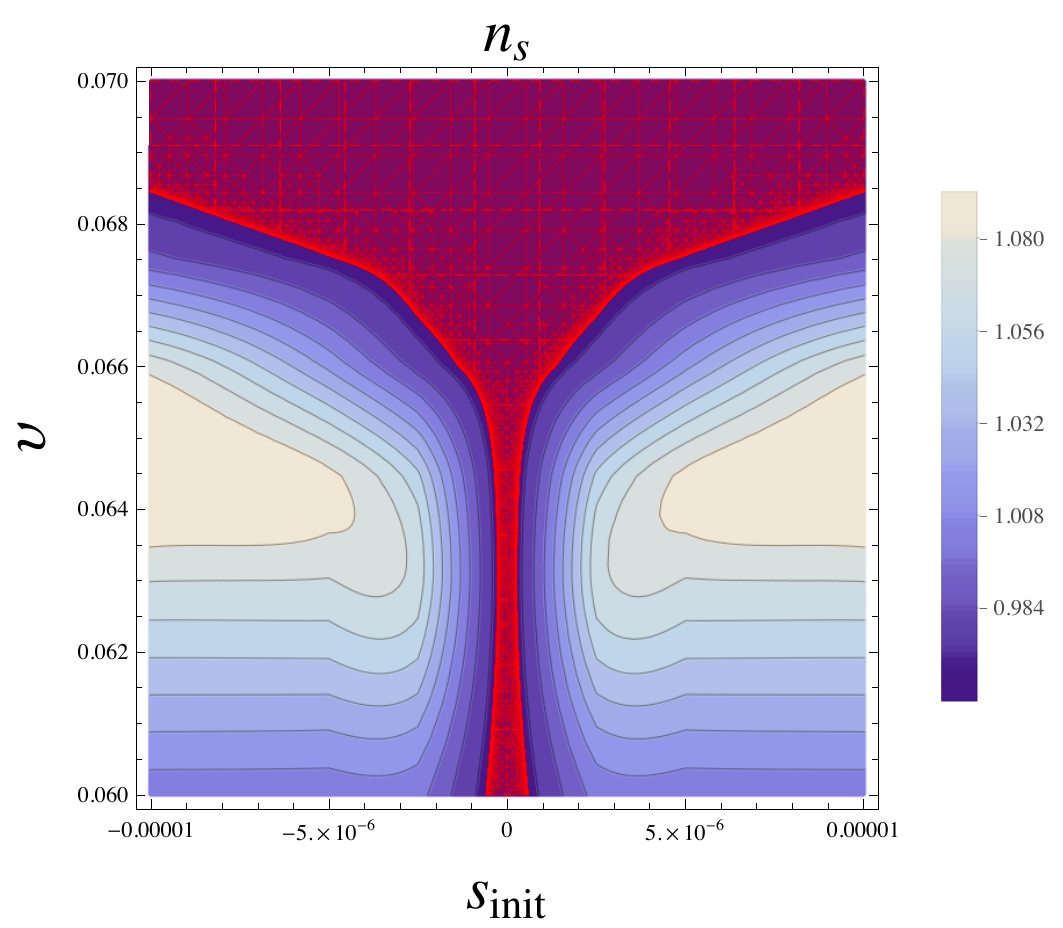}
%\includegraphics[width=89mm]{Fignscont_r.eps}
%\end{array}
%\end{eqnarray*}
\caption{\label{fig:ns}
The behavior of the scalar spectral index $n_s$ as $\upsilon$ and $s_{\rm init}$ are varied.
The left panel shows the $n_s$ as a function of $\upsilon$ 
($s_{\rm init}=0$, $1.0\times 10^{-7}$, $1.0\times 10^{-6}$, and $1.0\times 10^{-5}$),
and the right panel is a contour plot in the $s_{\rm init}$--$\upsilon$ plane.
The initial conditions for $\dot s_{\rm init}$, $h_{\rm init}$, and $\dot h_{\rm init}$ are the same as in Fig. \ref{fig:PotView}.
The green-shaded region on the left panel and the red-shaded region on the right panel indicate 
the 68\% C.L. constraints by Planck \cite{Ade:2013zuv}.
}
\end{figure*}
%\end{center}
%%%%%%%%%%%%%%%%%%%%%%%%%%%%%%%%%%%%%%%%%%

\subsubsection{Tensor power spectrum}

Figure~\ref{fig:At} shows the behavior of the amplitude of the tensor mode fluctuations $A_t$.
We assume this to be normalized at the pivot scale $k_0=0.05$ Mpc~\cite{Ade:2013uln},
\begin{align}
{\mathcal P}_T=A_t \left(\frac{k}{k_0}\right)^{n_t+\half\frac{d n_t}{d\ln k}\ln\frac{k}{k_0}+\cdots}.
\end{align}
The change of the initial value $s_{\rm init}$ and the shift of $\upsilon$ have no effects on $A_t$.
This is expected, since the tensor mode fluctuations are generated inside the horizon and do not interact once they exit the horizon.

Thus, the behavior of the tensor-to-scalar ratio $r$ is entirely determined by that of the scalar
amplitude $A_s$ (Fig.~\ref{fig:r}).
Consequently, the multifield effects only lower the tensor-to-scalar ratio $r$.

\subsubsection{Scalar spectral index}

In Fig.~\ref{fig:ns}, we show the behavior of the scalar spectral index $n_{s}$.
The behavior agrees well with our understanding that the curvature perturbation is sourced by
the isocurvature perturbation at superhorizon scales, and this effect shifts the spectral index.
Comparing with Fig.~\ref{fig:fNL}, we see that the Planck constraints on $f_{\rm NL}$ impose a more stringent bound on $\upsilon$ than $n_s$.
The contour plot on the right panel shares some similarity with the $f_{\rm NL}$ case;
this is attributed to the fact that the shift of $n_s$ is controlled by the derivative of the transfer function $T_{{\zeta}{\C S}}$ \cite{Peterson:2010mv,Kaiser:2012ak}.

\subsubsection{Tensor spectral index}

\begin{figure}[b]
%\begin{eqnarray*}
%\begin{array}{ccc}
\includegraphics[width=87mm]{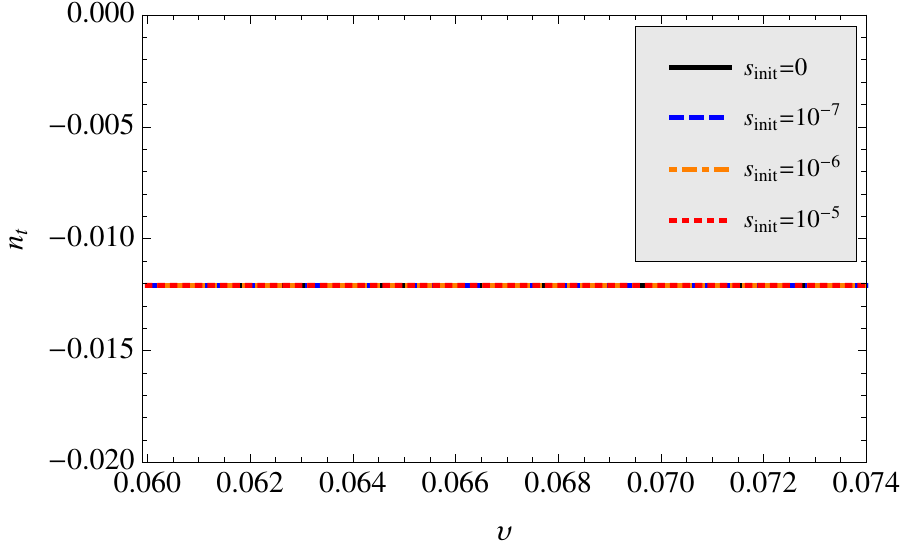}
%\includegraphics[width=87mm]{Fignt.eps}
%\end{eqnarray*}
\caption{\label{fig:nt}
The tensor spectral index $n_t$.
The initial conditions for $\dot s_{\rm init}$, $h_{\rm init}$, and $\dot h_{\rm init}$ are the same as in Fig. \ref{fig:PotView}.
}
\end{figure}

Finally, we show the tensor spectral index $n_t$ in Fig. \ref{fig:nt}.
As the tensor fluctuations do not interact with the scalar mode and hence freeze once they
exit the horizon, it is insensitive to the change of the inflaton trajectory.
Thus, the value of $n_t$ does not depend on $\upsilon$ nor on $s_{\rm init}$.

%%%%%%%%%%%%%%%%%%%%%%%%%%%%%%%%%%%%%%%%%%%%
\subsection{Constraints on the K\"{a}hler potential from non-Gaussianity}

%%%%%%%%%%%%%%%%%%%%%%%%%%%%%%%%%%%%%%%%%%%%
%%%%%%%%%%%%%%%%%%%%%%%%%%%%%%%%%%%%%%%%%%%%

% Summary of reuslts
We studied above the cosmological observables of the two-field HLI model in the phenomenologically interesting case of the seesaw mass scale $M=1\,\rm{TeV}$ and the $e$-folding number $N_e=60$.
For the initial values of the $s$ field $|s_{\rm init}|\lesssim 10^{-5}$ that are naturally expected from quantum fluctuations, we have seen that the nonlinearity parameter $f_{\rm NL}$, then the
scalar spectral index and the scalar power spectrum deviate from the observationally supported 
single-field values as we vary the K\"{a}hler potential parameter $\upsilon$ from above.
Within the range of the parameters we have searched, we did not see significant change in the tensor power spectrum and the tensor spectral index.
Putting the value of the tensor-to-scalar ratio $r$ aside, the observational constrains for the
scalar spectral index $n_s$ and the local-type nonlinearity parameter $f_{\rm NL}$ thus give constraints on the $\upsilon$ parameter.
Figures \ref{fig:As}, \ref{fig:fNL}, and \ref{fig:ns} show that among them the nonlinearity parameter puts the most stringent bound, $\upsilon\gtrsim 0.06925$.
The large nonlinearity parameter at small values of $\upsilon$ is understood as a consequence of the nontrivial inflaton trajectory.
Generation of non-Gaussianities involves several competing effects.
It is known that large non-Gaussianities can be generated when
the trajectory makes a turn after spending sufficient $e$-foldings on an unstable potential.
In fact, there are many studies in the literature (especially before the Planck satellite mission) in search of an inflationary model generating large non-Gaussianities.
While non-Gaussianities in the primordial fluctuations are still elusive, our case study above shows
that the observational bounds are useful in constraining the model parameters.

% Other parameters
We have also analyzed the model for various other parameter values.  
For example, in the case of $M=10\,\rm{TeV}$ and the same value of the $e$-folding $N_e=60$,
we have $\xi\approx 0.04612$ from the Planck normalization of the curvature perturbation 
in the single-field limit.
We obtained similar constraints on the $\upsilon$ parameter:
the $68\%$ bounds on the scalar spectral index giving $\upsilon \gtrsim 0.01257$ and
the $68\%$ bounds on $f_{\rm NL}$ giving $\upsilon \gtrsim 0.01246$.
The scalar power spectrum changes at a smaller value of $\upsilon$, and the tensor power spectrum and the tensor spectral index barely change.
Thus, we conclude $\upsilon \gtrsim 0.01257$, the strongest bound put by the scalar spectral index.
We find similar features in other parameter values.

%%%%%%%%%%%%%%%%%%%%%%%%%%%%%%%%%%%%%%%%%%%%
%%%%%%%%%%%%%%%%%%%%%%%%%%%%%%%%%%%%%%%%%%%%

%%%%%%%%%%%%%%%%%%%%%%%%%%%%%%%%%%%%%%%%%%
%\begin{center}
%\end{center}
%%%%%%%%%%%%%%%%%%%%%%%%%%%%%%%%%%%%%%%%%%

%%%%%%%%%%%%%%%%%%%%%%%%%%%%%%%%%%%%%%%%%%%%
%%%%%%%%%%%%%%%%%%%%%%%%%%%%%%%%%%%%%%%%%%%%
\section{Conclusion}\label{sec:concl}

In this paper, we analyzed the multifield dynamics of the supersymmetric Higgs inflation-type cosmological models that are implemented in (beyond) the Standard Model embedded in 
supergravity with a noncanonical K\"{a}hler potential. 
We studied the two-field HLI model based on the supersymmetric seesaw model in detail and studied how the cosmological observables constrain the model parameters.
Realistic particle theory-based inflationary models commonly involve multiple flat directions, and in such models, the understanding of multidimensional inflaton dynamics is crucial.
In the recent Planck satellite experiment, no sign of non-Gaussianities was detected \cite{Ade:2013ydc}.
However, as we have illustrated in this paper, the upper bound of the nonlinearity parameter
\eqref{eqn:planck} can be useful for obtaining information on the inflaton trajectory; in our case, this led to the constraints on the K\"{a}hler potential.

The main focus of the present paper was the local-type nonlinearity parameter $f_{\rm NL}$, that is,
the bispectrum of the primordial density fluctuations.
While $f_{\rm NL}$ is relatively a clean signal of multifield inflation, it is certainly not the only parameter that characterizes multifield inflation. 
While more dependent on details of the phenomenological setup, constraints from the isocurvature modes are also a rich source of information.
In models of inflation in which the neutrino sector is involved (such as the HLI model), the baryon number density through possible leptogenesis may also provide further constraints on the model (as discussed in the line of Refs. \cite{Murayama:1992ua,Murayama:1993xu}).
Furthermore, higher-order non-Gaussianities such as the trispectrum 
%($g_{\rm NL}$ and $\tau_{\rm NL}$) 
will naturally give richer information on the multifield dynamics, although, regarding the present status of these parameters \cite{Ade:2013ydc}, 
reliable constraints may not be obtainable in the near future.

%%%%%%%%%%%%%%%%%%%%%%%%%%%%%%%%%%%%%%%%%%%%
%%%%%%%%%%%%%%%%%%%%%%%%%%%%%%%%%%%%%%%%%%%%

%%%%%%%%%%%%%%%%%%%%%%%%%%%%%%%%%%%%%%%%%%%%
%%%%%%%%%%%%%%%%%%%%%%%%%%%%%%%%%%%%%%%%%%%%
%{\em Acknowledgements.}---
\subsection*{Acknowledgments}
We thank Nobuchika Okada for helpful discussions and comments.
We also acknowledge helpful conversations with Chris Byrnes, Ki-Young Choi, 
Misao Sasaki, and Takahiro Tanaka.
This research was supported in part by the National Research Foundation of Korea Grant-in-Aid for Scientific Research
NRF-2012R1A1A2007575
%(No. 2012007575?) 
(S.K.) and the NRF Global Ph.D. Fellowship Program No. 2011-0008792 (J.K.).
We used computing resources of the Yukawa Institute, Kyoto University.
\bigskip
%%%%%%%%%%%%%%%%%%%%%%%%%%%%%%%%%%%%%%%%%%%%
%%%%%%%%%%%%%%%%%%%%%%%%%%%%%%%%%%%%%%%%%%%%

%%%%%%%%%%%%%%%%%%%%%%%%%%%%%%%%%%%%%%%%%%%%
%%%%%%%%%%%%%%%%%%%%%%%%%%%%%%%%%%%%%%%%%%%%
\appendix
%%%%%%%%%%%%%%%%%%%%%%%%%%%%%%%%%%%%%%%%%%%%
%%%%%%%%%%%%%%%%%%%%%%%%%%%%%%%%%%%%%%%%%%%%

%%%%%%%%%%%%%%%%%%%%%%%%%%%%%%%%%%%%%%%%%%%%
%%%%%%%%%%%%%%%%%%%%%%%%%%%%%%%%%%%%%%%%%%%%

\section{Expressions for coefficients in the backward $\delta N$ formalism}\label{apdx:deltaN}

%%%%%%%%%%%%%%%%%%%%%%%%%%%%%%%%%%%%%%%%%%%%
%%%%%%%%%%%%%%%%%%%%%%%%%%%%%%%%%%%%%%%%%%%%
\subsection{$P^{a}_{b}$, $Q^{a}_{bc}$, and $A^{ab}$}\label{apdx:expressions}
The expressions for $P^{a}_{b}$ and $Q^{a}_{bc}$ can be obtained by performing the derivatives of the background equations of motion with respect to $\lambda$. The results are as follows:
\begin{align}
P^{I1}_{1J}&=0\,,\nonumber\qquad
P^{I2}_{1J}=\delta^{I}_{J}\,,\nonumber\\
P^{I1}_{2J}&=-\frac{\nabla_{J}\nabla^{I}V}{H^{2}}
+\frac{(\nabla^{I}V)(\nabla_{J}V)}{H^{2}V}
-R^{I}_{KJL}\varphi_{2}^{K}\varphi_{2}^{L}\,,\nonumber\\
P^{I2}_{2J}&=G_{JK}\varphi_{2}^{K}\varphi_{2}^{I}-\frac{V}{H^{2}}\delta^{I}_{J}
+\frac{\nabla^{I}V}{V}G_{JK}\varphi_{2}^{K}\,,
\end{align}
and
\begin{align}
Q^{I11}_{1JK}&=-R^{I}_{JKL}\varphi_{2}^{L}\,,\nonumber\\
Q^{I12}_{1JK}&=Q^{I21}_{1JK}=Q^{I22}_{1JK}=0\,,\nonumber\\
Q^{I11}_{2JK}&=-\left(\nabla_{J}R^{I}_{MKL}\right)\varphi_{2}^{M}\varphi_{2}^{L}
-\frac{\nabla_{K}\nabla_{J}\nabla^{I}V}{H^{2}}\nonumber\\
&\hspace{5mm}+\frac{2\left(\nabla_{J}\nabla^{I}V\right)\left(\nabla_{K}V\right)}{H^{2}V}
+\frac{\left(\nabla^{I}V\right)\left(\nabla_{K}\nabla_{J}V\right)}{H^{2}V}\nonumber\\
&\hspace{5mm}-\frac{2\left(\nabla^{I}V\right)\left(\nabla_{J}V\right)\left(\nabla_{K}V\right)}{H^{2}V^{2}}
\,,\nonumber\\
Q^{I12}_{2JK}&=-2R^{I}_{KJL}\varphi_{2}^{L}
+\frac{\nabla_{J}\nabla^{I}V}{V}G_{KL}\varphi_{2}^{L}\nonumber\\
&\hspace{5mm}-\frac{\left(\nabla^{I}V\right)\left(\nabla_{J}V\right)}{V^{2}}G_{KL}\varphi_{2}^{L}
\,,\nonumber\\
Q^{I21}_{2JK}&=-R^{I}_{LKJ}\varphi_{2}^{L}
+\frac{\nabla_{K}\nabla^{I}V}{V}G_{JL}\varphi_{2}^{L}\nonumber\\
&\hspace{5mm}-\frac{\left(\nabla^{I}V\right)\left(\nabla_{K}V\right)}{V^{2}}G_{JL}\varphi_{2}^{L}
\,,\nonumber\\
Q^{I22}_{2JK}&=2G_{JL}\varphi_{2}^{L}\delta^{I}_{K}
+G_{JK}\varphi_{2}^{I}
+G_{JK}\frac{\nabla^{I}V}{V}
\,.
\end{align}

The quantity $A^{ab}$ is defined via the two-point correlation function of field perturbations, $\delta\varphi^{a}$, as follows:
\begin{align}
\langle \delta\varphi^{a}_{*}\delta\varphi^{b}_{*}\rangle = A^{ab}\left(\frac{H_{*}}{2\pi}\right)^{2}\,.
\end{align}
After including the slow-roll corrections, the $(1,1)$ component of $A^{ab}$ is given by \cite{Nakamura:1996da}
\begin{align}
\langle \delta\varphi_{1*}^{I}\delta\varphi_{1*}^{J}\rangle
=\left(\frac{H_{*}}{2\pi}\right)^{2}\left[
G^{IJ}-2\epsilon G^{IJ}+2\alpha\, \epsilon_{KL} M^{KL}G^{IJ}
\right]_{*}\,,
\end{align}
where
\begin{align}
\epsilon_{KL}&\equiv\epsilon\, G_{KL}+\left(
G_{KM}G_{LN}-\frac{1}{3}R_{KMLN}
\right)\varphi_{2}^{M}\varphi_{2}^{N}
\nonumber\\
&\hspace{5mm}-\frac{\nabla_{K}\nabla_{L}V}{3H^{2}}\,,
\end{align}
and
\begin{align}
M_{KL}\equiv \frac{N_{K}^{1}N_{L}^{1}}{G^{AB}N_{A}^{1}N_{B}^{1}}\,.
\end{align}

The other components of $A^{ab}$ can be obtained by considering the derivatives of the background equations of motion with respect to $\lambda$. Assuming that the slow-roll conditions are satisfied at the Hubble exit, i.e., at $N=N_{*}$, we find
\begin{align}
\overset{(1)}{\delta\varphi_{2}^{I}}\approx
\left[
\frac{\left(\nabla^{I}V\right)\left(\nabla_{J}V\right)}{V^{2}}
-\frac{\nabla_{J}\nabla^{I}V}{V}
\right]\overset{(1)}{\delta\varphi_{1}^{J}}
\equiv\Delta^{I}_{J}\overset{(1)}{\delta\varphi_{1}^{J}}\,.
\end{align}
The resultant expressions for the components of $A^{ab}$, with slow-roll corrections, are found to be
\begin{align}
A_{11}^{IJ}&=G^{IJ}-2\epsilon\, G^{IJ}+2\alpha\,G^{IJ}\epsilon_{KL}
\frac{G^{KC}G^{LD}N_{C}^{1}N_{D}^{1}}{G^{AB}N_{A}^{1}N_{B}^{1}} \,,\nonumber\\
A_{12}^{IJ}&=\Delta_{K}^{I}A_{11}^{KJ}\,,\qquad
A_{21}^{IJ}=(A_{12}^{IJ})^{T}\,,\nonumber\\
A_{22}^{IJ}&=\Delta_{K}^{I}\Delta_{L}^{J}A_{11}^{KL}\,,
\end{align}
where $T$ stands for the transpose.

%%%%%%%%%%%%%%%%%%%%%%%%%%%%%%%%%%%%%%%%%%%%
%%%%%%%%%%%%%%%%%%%%%%%%%%%%%%%%%%%%%%%%%%%%

%%%%%%%%%%%%%%%%%%%%%%%%%%%%%%%%%%%%%%%%%%%%
%%%%%%%%%%%%%%%%%%%%%%%%%%%%%%%%%%%%%%%%%%%%
\subsection{$N^{F}_{a}$ and $N^{F}_{ab}$}\label{apdx:inicond}
The explicit expressions of $N^{F}_{a}$ and $N^{F}_{ab}$ are as follows:
\begin{align}
N_{I}^{1}&=\frac{\nabla_{I}V}{2\epsilon V}
\,,\quad
N_{I}^{2}=\frac{G_{IJ}\varphi_{2}^{J}}{2\epsilon(3-\epsilon)}
\,,
\end{align}
and
\begin{align}
N_{IJ}^{11}&=\frac{\nabla_{J}\nabla_{I}V}{2\epsilon V}
-\frac{\left(\nabla_{I}V\right)\left(\nabla_{J}V\right)}{\epsilon V^{2}}
+\frac{3\left(\nabla_{I}V\right)\left(\nabla_{J}V\right)}{2\epsilon^{2}V^{2}}
\nonumber\\
&\hspace{5mm}+\frac{3-\epsilon}{4\epsilon^{3}V^{3}}\left(\nabla_{I}V\right)\left(\nabla_{J}V\right)\left(\nabla_{K}V\right)\varphi_{2}^{K}
\,,\nonumber\\
N_{IJ}^{12}&=\frac{3\left(\nabla_{I}V\right)G_{JK}\varphi_{2}^{K}}{2\epsilon^{2}(3-\epsilon)V}
+\frac{G_{JL}\varphi_{2}^{L}}{4\epsilon^{3}V^{2}}\left(\nabla_{I}V\right)\left(\nabla_{K}V\right)\varphi_{2}^{K}
\nonumber\\
&\hspace{5mm}-\frac{\left(\nabla_{I}V\right)G_{JK}\varphi_{2}^{K}}{2\epsilon^{2}V}
-\frac{\left(\nabla_{I}V\right)G_{JK}\varphi_{2}^{K}}{2\epsilon(3-\epsilon)V}
\,,\nonumber\\
N_{IJ}^{21}&=\frac{3\left(\nabla_{J}V\right)G_{IK}\varphi_{2}^{K}}{2\epsilon^{2}(3-\epsilon)V}
+\frac{G_{IL}\varphi_{2}^{L}}{4\epsilon^{3}V^{2}}\left(\nabla_{J}V\right)\left(\nabla_{K}V\right)\varphi_{2}^{K}
\nonumber\\
&\hspace{5mm}-\frac{\left(\nabla_{J}V\right)G_{IK}\varphi_{2}^{K}}{2\epsilon^{2}V}
-\frac{\left(\nabla_{J}V\right)G_{IK}\varphi_{2}^{K}}{2\epsilon(3-\epsilon)V}
\,,\nonumber\\
N_{IJ}^{22}&=\frac{G_{IJ}}{2\epsilon(3-\epsilon)}
+\frac{3G_{IK}G_{JL}\varphi_{2}^{K}\varphi_{2}^{L}}{2\epsilon^{2}(3-\epsilon)^{2}}
\nonumber\\
&\hspace{5mm}
+\frac{G_{IK}G_{JL}\varphi_{2}^{K}\varphi_{2}^{L}}{4\epsilon^{3}(3-\epsilon)V}\left(\nabla_{M}V\right)\varphi_{2}^{M}
-\frac{G_{IK}G_{JL}\varphi_{2}^{K}\varphi_{2}^{L}}{\epsilon^{2}(3-\epsilon)}
\nonumber\\
&\hspace{5mm}
-\frac{G_{IK}G_{JL}\varphi_{2}^{K}\varphi_{2}^{L}}{2\epsilon(3-\epsilon)^{2}}
\,.
\end{align}
%
%%%%%%%%%%%%%%%%%%%%%%%%%%%%%%%%%%%%%%%%%%%%
%%%%%%%%%%%%%%%%%%%%%%%%%%%%%%%%%%%%%%%%%%%%

%%%%%%%%%%%%%%%%%%%%%%%%%%%%%%%%%%%%%%%%%%%%
%%%%%%%%%%%%%%%%%%%%%%%%%%%%%%%%%%%%%%%%%%%%

%%%%%%%%%%%%%%%%%%%%%%%%%%%%%%%%%%%%%%%%%%%%
%%%%%%%%%%%%%%%%%%%%%%%%%%%%%%%%%%%%%%%%%%%%
%%%%%
\section{NMSSM Higgs inflation}\label{apdx:NMSSM}
%%%%%
Higgs inflation based on the NMSSM is studied in Refs. \cite{Einhorn:2009bh,Ferrara:2010yw,Ferrara:2010in}.
The NMSSM is also an extension of the MSSM by a singlet field $S$, and the superpotential of
%This type of model has the superpotential
%\begin{align}
%W\sim SX^2+S^3.
%\end{align}
its simplest version (the ${\B Z}_3$ invariant NMSSM) is (see Ref. 
\cite{Ellwanger:2009dp} for a review)
\beq
W_{\rm NMSSM}=W_{\rm MSSM}+\lambda S H_u H_d+\frac{\rho}{3} S^3,
\label{eqn:WNMSSM}
\eeq
where $W_{\rm MSSM}$ is Eq. \eqref{eqn:WMSSM}.
Since one of the motivations for considering the NMSSM is to solve the MSSM $\mu$ problem by
generating an effective $\mu$ term as the expectation value of the $S$ field, the $\mu H_u H_d$ 
term of Eq. \eqref{eqn:WMSSM} is usually not included in the NMSSM.
The parameters $\lambda$ and $\rho$ in Eq. \eqref{eqn:WNMSSM} are not arbitrary but are constrained by the conditions that (i) $\langle S\rangle=0$ is not the global minimum (leading to $\rho^2<\lambda^2$), 
(ii) there is no Landau singularity below the grand unified theory scale (leading to $\lambda\lesssim 0.8$), and
(iii) $\mu_{\rm eff}$ is in the electroweak scale ($\lambda$ not too small).
If one gives up solving the MSSM $\mu$ problem and keeps the $\mu H_u H_d$ term of the 
MSSM, then $\lambda$ can be taken arbitrarily small, and the NMSSM approaches the MSSM limit.
In the NMSSM Higgs inflation model, the nonminimal coupling $\xi=\frac{\gamma}{4}-\sixth$ 
(see also the main text) is related to $\lambda$
via the normalization of the curvature fluctuations.
The value of $\xi$ can be ${\C O}(1)$ if $\lambda$ is allowed to be tuned small.

The NMSSM Higgs inflation model \cite{Einhorn:2009bh,Ferrara:2010yw,Ferrara:2010in} is obtained
by supergravity embedding with the K\"{a}hler potential $K\equiv -3\Phi$ (we use the superconformal framework) with
\begin{align}
\Phi=&1-\third\left(|S|^2+|H_u^0|^2+|H_d^0|^2\cdots\right)\nn\\
&\qquad+\frac{\gamma}{2}\left(H_u^0 H_d^0+c.c.\right)+\frac{\upsilon}{3}|S|^4.
\end{align}
The ellipsis represents the canonical K\"{a}hler terms for the other fields that are not relevant to the
study of inflation.
Setting the charged Higgs to be zero and assuming that the $H_u$-$H_d$ D-flat direction and the singlet direction are parametrized by two real scalar fields $h$, $s$ as
\beq
H_u^0=\frac{h}{2},
\quad
H_d^0=\frac{h}{2},
\quad
S=\frac{s}{\sqrt 2},
\label{eqn:hands}
\eeq
the scalar-gravity part of the Lagrangian becomes Eq. \eqref{eqn:Lag} in the Einstein frame.
The scalar potential is $V(\phi^I)=\Phi^{-2}V_{\rm J}(s,h)$, 
with the Jordan frame potential in this case 
reading
\begin{align}\label{eqn:NMSSMVJ}
V_{\rm J}(s,h)
&=\quarter \lambda^2 s^2 h^2
+\frac{\left(\frac{\lambda}{2}h^2+\rho s^2\right)^2}{4(1-2\upsilon s^2)}
\nn\\
&-\frac{3}{32}\frac{\left\{\lambda\gamma s h^2-\frac 23 \frac{\upsilon s^3}{1-2\upsilon s^2}
\left(\frac{\lambda}{2} h^2+\rho s^2\right)\right\}^2}
{1+\frac{\gamma}{4}\left(\frac 32 \gamma-1\right) h^2
+\frac{\upsilon s^4}{12(1-2\upsilon s^2)}}.
\end{align}
There is a typo in Eq. (D4) of Ref. \cite{Ferrara:2010in}.
The function $\Phi$ takes the form of Eq. \eqref{eqn:Phi}.
While Eq. \eqref{eqn:NMSSMVJ} differs from Eq. \eqref{eqn:VJ} in details, the overall shape of the potential 
is similar.
In the Einstein frame, the K\"{a}hler potential parameter $\upsilon$ controls the stability of the potential $V(\phi^I)$ in the $s$ direction.
Decreasing the value of $\upsilon$ enhances the multifield effects, including the non-Gaussianities of the primordial fluctuations.
It is thus possible to constrain the value of $\upsilon$ using the observational bounds of non-Gaussianities, as we did in this paper for the HLI model.

%%%%%%%%%%%%%%%%%%%%%%%%%%%%%%%%%%%%%%%%%%%%
%%%%%%%%%%%%%%%%%%%%%%%%%%%%%%%%%%%%%%%%%%%%

%%%%%%%%%%%%%%%%%%%%%%%%%%%%%%%%%%%%%%%%%%%%
%%%%%%%%%%%%%%%%%%%%%%%%%%%%%%%%%%%%%%%%%%%%
\end{document}